\documentclass[aps,prb,twocolumn,superscriptaddress]{revtex4-2}
\usepackage{mathtools}
\usepackage{bm}
\usepackage{dsfont,amsthm,amsbsy}
\usepackage{verbatim}
\usepackage{amssymb} 
\usepackage{amsmath}
\usepackage{bbm}
\usepackage{graphicx}
\usepackage{epstopdf}
\usepackage{subfigure}
\usepackage{natbib}
\usepackage{epsfig}
\usepackage{amsfonts}
\usepackage{mathrsfs}
\usepackage{sidecap}
\usepackage{lipsum}
\usepackage[toc,page,title,titletoc,header]{appendix}
\usepackage[colorlinks,linkcolor=blue,citecolor=blue,anchorcolor=blue,urlcolor=blue]{hyperref}
\usepackage{hyperref}
\usepackage{resizegather}
\usepackage{tikz}
\usepackage{float}
\usepackage{mathbbol}
\usepackage[normalem]{ulem}
\usepackage{cancel}
\usepackage{upgreek}
\graphicspath{{Figures/}}

\newcommand{\be}{\begin{equation}}
\newcommand{\ee}{\end{equation}}
\newcommand{\bea}{\begin{eqnarray}}
\newcommand{\eea}{\end{eqnarray}}

\newcommand{\up}{\uparrow}
\newcommand{\down}{\downarrow}

\begin{document}
	
\title
{Dynamical Phase diagram of the Quantum Ising model with Cluster Interaction Under Noisy and Noiseless Driven field}
	
\author{Sasan Kheiri}
\affiliation{Department of Physics, Institute for Advanced Studies in Basic Sciences (IASBS), Zanjan 45137-66731, Iran}
\affiliation{Department of Physics, University of Guilan, 41335-1914 Rasht, Iran}

\author{R. Jafari}
\email[]{raadmehr.jafari@gmail.com}
\affiliation{Physics Department and Research Center OPTIMAS, University of Kaiserslautern, 67663 Kaiserslautern, Germany}
	
\author{S. Mahdavifar}
\affiliation{Department of Physics, University of Guilan, 41335-1914 Rasht, Iran}
	
\author{Ehsan Nedaaee Oskoee}
\affiliation{Department of Physics, Institute for Advanced Studies in Basic Sciences (IASBS), Zanjan 45137-66731, Iran}

\author{Alireza  Akbari}
\affiliation{Beijing Institute of Mathematical Sciences and Applications (BIMSA), Huairou District, Beijing 101408, P. R. China}

\date{\today}
	
\begin{abstract}
In most lattice models, gap closing typically occurs at high-symmetry points in the Brillouin
zone. In the transverse field Ising model with cluster interaction, besides the gap closing at high-symmetry points, 
the gap closing at the quantum phase transition between paramagnetic and cluster phases of the model can be moved by 
tuning the strength of the cluster interaction. We take advantage of this property to examine the nonequilibrium dynamics 
of the model in the framework of dynamical quantum phase transitions (DQPTs) after a noiseless and noisy ramp of 
the transverse magnetic field. The numerical results show that DQPTs always happen if the starting or ending point of the
quench field is restricted between two critical points. In other ways, there is always critical 
sweep velocity above which DQPTs disappear. Our finding reveals that noise modifies drastically the dynamical phase diagram of 
the model. We find that the critical sweep velocity decreases by enhancing the 
noise intensity and scales linearly with the square of noise intensity for weak and strong noise. Moreover, the region 
with multi-critical modes induced in the dynamical phase diagram by noise. The sweep velocity 
under which the system enters the multi-critical modes (MCMs) region increases by enhancing the noise and scales
linearly with the square of noise intensity.  
\end{abstract}

\maketitle

	
\section{Introduction}

Developing a comprehensive theoretical framework for nonequilibrium phenomena is a challenging 
problem in physics with an impact vastly surpassing this specific discipline \cite{Hohenberg1997,Chou2011,Mishra2013,Chanda2016,Awasthi2018}.  
A systematic understanding is crucial for many different areas such as many-body correlations \cite{Abanin2019,Nandkishore2015}, 
quantum matters \cite{Wilczek2012,Goldman2014}, quantum simulations \cite{Erne2018,Bojan2018}, 
and quantum technologies, \cite{Albash2018,Georgescu2014}, which require the control of many-body physical systems at the quantum level. 
This question has recently promoted experimental and theoretical studies of the out-of-equilibrium dynamics of many-body systems \cite{Polkovnikov2011,Lamporesi2013,Schneider2012}. 

Recent experimental advances on synthesizing various quantum platforms, including ultra-cold atoms in optical lattices \cite{Jotzu2014,Daley2012,Schreiber2015,Choi2016,Flaschner2018}, trapped ions \cite{Jurcevic2017,Martinez2016,Neyenhuis2017,Smith2016}, 
nitrogen-vacancy centers in diamond \cite{Yang2019}, superconducting qubit systems \cite{Guo2019} and quantum walks in photonic systems \cite{Wang2019,Xu2020} 
provide a framework for experimentally studying the out-of-equilibrium dynamics of many-body systems.

Nevertheless, the simulation of the adapted time dependent Hamiltonian in any real experiment is imperfect and noisy fluctuations 
are imminent. In other words, the noises are ubiquitous and imperative in any physical system and the stability of 
a dynamical system can be strongly affected in the presence of uncontrolled perturbations such as random noise \cite{Horsthemke1984,Zoller1981,Chen2010,Doria2011}.

Consequently, understanding the effect of noise on the Hamiltonian evolution is crucial for correctly
predicting the results of experiments and designing experimental setup robust against the effect of the noise \cite{Pichler2012}.

Within this context, we investigate the effects of Gaussian white noise on the dynamical quantum phase transition (DQPTs). 
Dynamical quantum phase transitions have become one of the focal points in the study of quantum matter out of equilibrium \cite{Heyl2013,Heyl2017,Heyl2018}.
DQPT proposed theoretically in nonequilibrium quantum systems inspired by the concept of nonanalyticities associated with the 
free-energy density of a classical system at a finite temperature transition. The DQPT is signaled through the nonanalytical 
behavior of dynamical free energy,\cite{Andraschko2014,Karrasch2013,Jafari2019a,Mondal2022,Mendoza2022,
Sedlmayr2018a,Sedlmayr2018b,Khatun2019,Ding2020,Rossi2022,Khan2023,Vajna2014,Porta2020}, where the real-time plays the role of the control parameter \cite{Jafari2019b,Jafari2017,Najafi2019,Sadrzadeh2021,Wong2022,Rylands2021,Abdi2019,Uhrich2020,Wong2023,Sacramento2024}. 
DQPT displays a phase transitions between dynamically emerging quantum phases, that takes place during the nonequilibrium
coherent quantum time evolution under sudden quench and ramp protocols \cite{Zhou2021,Vanhala2023,Mondal2023,Cao2020,Sedlmayr2023,Sedlmayr2022,Sedlmayr2020,
Zeng2023,Stumper2022,Yu2021,Vijayan2023,Xue2023,Bhattacharjee2023,Leela2022,Zamani_2024,Mishra2020,Haldar2020,Jad2021,Jad2024,Jad2023,Jad2021b} or time-periodic
modulation of Hamiltonian \cite{Yang2019,Zamani2020,Kosior2018a,Jafari2021,Kosior2018b,Naji2022,Jafari2022,Naji2022b}.

It has been also established that there exists a dynamical topological order parameter (DTOP), similar to the order parameters 
at conventional quantum phase transition, which uncover DQPTs \cite{Budich2016,Bhattacharya2017}. The presence of DTOP, which takes 
integer values as a function of time and jumps at the critical times, represents the emergence of a topological characteristic 
associated with the time evolution of nonequilibrium systems. Theoretical predictions of DQPT were confirmed experimentally in several studies \cite{Flaschner2018,Jurcevic2017,Martinez2016,Guo2019,Wang2019,Nie2020,Tian2020}.
Most of the research dedicated to deterministic quantum evolution induces by sudden quench or ramp of the Hamiltonian parameters. 
However, inadequate consideration has been associated with the stochastic driving of thermally isolated systems with noisy Hamiltonian \cite{Jafari2024,Baghran2024,Rahmani2016,Bando2020,Chenu2017}.

In this work we study DQPTs in the one-dimensional quantum Ising model with cluster interaction (three site spin) \cite{Verga2023,Son2011,Smacchia2011,Niu2012,Montes2012,Kopp2005,Hirsch1979,Doherty2009,Yu2024,Ding2019} 
in the presence of the noiseless and noisy linear driven transverse field. The cluster interaction is generated 
in the first step of a real space renormalization group procedure \cite{Hirsch1979}.
In most lattice models, the gap closing typically occurs at the edges or center of the Brillouin zone (high-symmetry points). 
In the time-independent transverse field Ising model with cluster interactions, beside the gap closing at 
the high-symmetry points in the Brillouin zone, the gap closing occurs at the quantum phase transition between the paramagnetic 
and cluster phases of the model which can be moved by tuning the cluster interaction strength.

Moreover, the cluster interaction breaks the usual symmetry of the transverse field Ising model phase diagram.
We take advantage of this property to explore the nonequilibrium dynamics of the model using the notion of DQPTs
following the noiseless and noisy ramping of the transverse field. The phase diagram symmetry breaking is dramatic, 
as the DQPTs features are distinct for ramp down and ramp up of the transverse magnetic field.
%
\begin{figure}[t]
\centering
{\includegraphics[width=\linewidth]{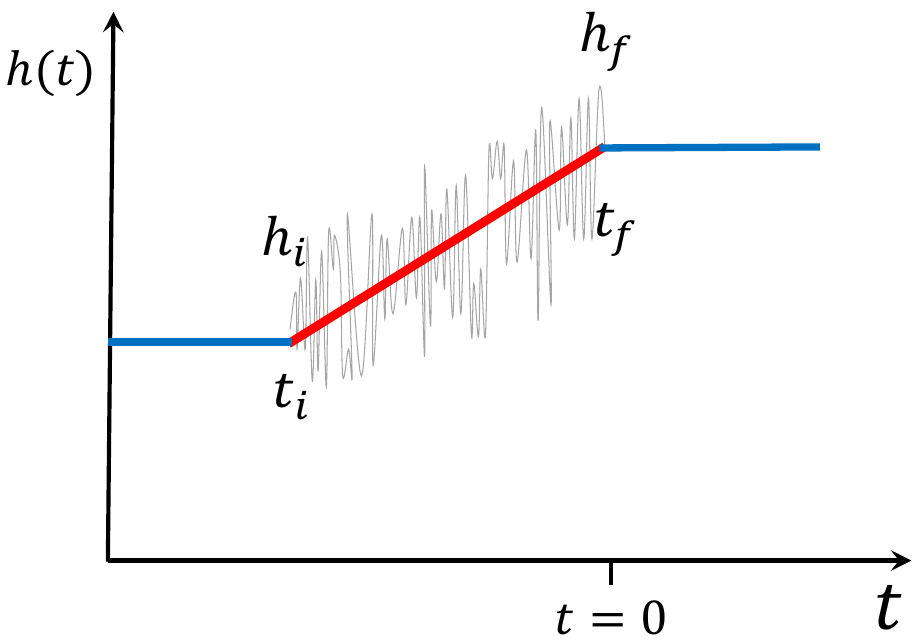}}
\caption{ A schematic representation of a linear ramp protocol accompanied 
by noise fluctuations. The quenching process starts at $t_{i}< 0$ with the 
magnetic field $h(t)$ set to $h_i$ and ends at $t_f$, where $h(t_{f}) = h_f$.}
\label{fig1}
\end{figure} 
%

We show that, when the ramp filed crosses the critical points (does not matter how many QCPs are crossed) there exist a critical sweep velocity 
above which DQPTs are whipped out, if the starting or ending point of the ramp field is not confined between two critical points.  
Otherwise, DQPTs always occur even for a sudden quench case. In addition, we discover that the sweep velocity above which DQPTs disappear 
decreases by enhancing the noise intensity and scales linearly with the square of noise intensity for both weak and strong noise, which support
earlier findings \cite{Jafari2025b}.

Further, noise induces multi-critical modes (MCMs) region (continuous critical time) in the dynamical phase diagram of the model. 
The numerical results show that the sweep velocity below which the system enters the MCMs region enhances by increasing the noise intensity and also 
scales linearly with the square of noise intensity. \\

The paper is organized as follows.
In Sec. \ref{section2}, the dynamical free energy and dynamical topological order parameter of the two level Hamiltonians are discussed.
In Section \ref{section3}, we introduce the model along with its exact solution and equilibrium phase diagram.
Section \ref{section4} is dedicated to the numerical simulation of the noiseless case based on the analytical result.
Section \ref{section5} focuses on the numerical simulation of the model utilizing the exact noise master equation.
Section \ref{section6} contains some concluding remarks.


\section{Ramp protocol in an integrable model}\label{section2}

\subsection{Dynamical free energy}
    
We adopt  the terminology used in Refs. \cite{Sharma2016,Zamani_2024,Jafari2024} for all ramp schemes, which will be examined in the discussions that follow.
Suppose we have an integrable model that can be reduced to a two-level Hamiltonian, denoted as $H_{k}(h)$, for each momentum mode $k$. At the initial 
time ($t_{i} \to - \infty$), the system is in the ground state $| \alpha^{i}_{k}\rangle $ of the pre-quench Hamiltonian $H_{k}(h_{i})$ for each mode.
In the ramp protocol, the Hamiltonian is described by a parameter $h$, varies from an initial value $h_i$ at time $t_i$, following the 
linear time driven scheme $h(t) = vt$, to a final value $h_f$ at time $t_f$ (Fig. \ref{fig1}). This is designed so that the system crosses the quantum critical point (QCP) at $h= h_c$.

Crossing the critical point (gap closing point) disrupts the adiabatic condition, leading to a non-adiabatic transition. 
Therefore the final state $| \psi_{k}(h_f) \rangle=| \psi_{k}^{f} \rangle $ (corresponding to the $k$-th mode) may not be the ground state of 
the post-quench Hamiltonian $H_{k}(h_f)=H_{k}^{f}$.

Consequently, the final state should be given as a linear combination of the ground and excited states $| \psi_{k}^{f} \rangle = a_{k} |\alpha_{k}^{f}\rangle +b_{k} | \beta_{k}^{f}\rangle $, ($|a_{k}|^{2}+|b_{k}|^{2}=1$) where, $|\alpha_{k}^{f}\rangle$ and $|\beta_{k}^{f} \rangle$ are the ground and the excited states of the post-quench Hamiltonian $H_{k}^{f}$, respectively with the corresponding energy eigenvalues $\epsilon^{f}_{k,1}$ and $\epsilon^{f}_{k,2}$.

The probability of non-adiabatic transition resulting in the system being in the excited state at $h=h_f$ is represented as 
$p_{k}=|b_{k}|^{2}=|\langle\beta_{k}^{f}|\alpha_{k}^{i}\rangle |^{2}$. Consequently, the Loschmidt overlap and the associated dynamical 
free energy \cite{Heyl2013,Heyl2018}, for mode $k$ at $t > t_f$ are specified by \cite{Sharma2016,Zamani_2024,Jafari2024}
%
\begin{eqnarray}
\nonumber
{\cal{G}}_{k} &=& \langle \psi^{f}_{k}|\exp({-iH^{f}_{k}t)} |\psi^{f}_{k} \rangle \\
\nonumber
&=& |a_{k}|^{2}\exp({-i\epsilon^{f}_{k,1}t)}+  |b_{k}|^{2}\exp({-i\epsilon^{f}_{k,2}t)}\\
\label{eq1}
g_{k}(t)  &=& -\frac{1}{N} \ln |{\cal{G}}_{k}|^{2}               
\end{eqnarray}
%
respectively, where $N$ is the size of the system.
		  
By summing the contributions from all modes and substituting the summation with an integral in the thermodynamic limit, one obtains
\cite{Sharma2016,Dora2013,Zamani_2024,Jafari2024}
%
{\small 
\begin{equation}
\label{r(t)}
g(t)=-\frac{1}{2 \pi} \int_{0}^{\pi}\ln\left(1+4p_{k}(p_{k}-1)\sin^{2}(\frac{\epsilon_{k,2}^{f}-\epsilon_{k,1}^{f}}{2})t\right) \, dk
\end{equation}}	    	
%
where the parameter $t$ is defined as the time elapsed since the final state, $|\psi_{k}^{f} \rangle$, is achieved at the end of the ramp process (Fig. \ref{fig1}). 
The non-analyticities in $g(t)$ occur at the values of the real time $t^{*}_{n}s$ given by
%
\begin{equation}
\label{tstar}
t^{*}_{n}=\frac{\pi}{\epsilon_{k^{*},2}^{f}-\epsilon_{k^{*},1}^{f}}(2n+1)
\end{equation}
%
These are the critical times for the DQPTs, with $k^{*}$ the mode at which the argument of the logarithm in Eq. \eqref{r(t)} vanishes for $|b_{k^{*}} |^{2} = p_{k^{*}} = 1/2$. For the case $\epsilon_{k^{*},2}^{f}=-\epsilon_{k^{*},1}^{f}=\epsilon_{k^{*}}^{f}$, Eq. \eqref{tstar} is simplified to
%
\begin{equation}
\label{t*}
t^{*}_{n}=t^{*}\left(n+\frac{1}{2}\right), \quad t^{*}=\frac{\pi}{\epsilon_{k^{*}}^{f}}
\end{equation}
%

\subsection{Dynamical Topological Order Parameter}

The dynamical topological order parameter is introduced to represent the topological characteristics associated with DQPTs \cite{Budich2016}. 
The DTOP exhibits quantized integer values over time, with unit magnitude jumps occurring at DQPTs, revealing the topological nature of DQPT\cite{Budich2016,Bhattacharjee2018}. The DTOP is derived from the gauge-invariant Pancharatnam geometric phase connected 
to the Loschmidt amplitude \cite{Budich2016,Bhattacharya2017}. The dynamical topological order parameter is defined as \cite{Budich2016}
%
\begin{eqnarray}
\label{eqDTOP}
N_{w}(t)=\frac{1}{2 \pi} \int_{0}^{\pi} {\frac{\partial \Theta^{G}(k,t)}{\partial k} dk},
\end{eqnarray}
%
where the geometric phase $\Theta^{G}(k,t)$ is obtained from the total phase $\Theta(k,t)$ through the subtraction of the dynamical phase 
$\Theta^{D}(k,t)$:  $ \Theta^{G}(k,t)=  \Theta(k,t)-  \Theta^{D}(k,t)$.
The total phase $\Theta(k,t)$ is the phase factor of Loschmidt amplitude in its polar coordinate representation, i.e., 
${\cal{G}}_{k}(t) = |{\cal{G}}_{k}(t)|\;e^{-i\Theta(k,t)}$, and
$\Theta^{D}(k,t)=\int_{0}^{t} \langle\psi_{k}^{f}(t')|H(k,t')|\psi_{k}^{f}(t')\rangle dt'$, in which $\Theta(k,t)$ and $\Theta^{D}(k,t)$, 
for the two level system can be calculated as follows \cite{Sharma2016,Zamani_2024,Jafari2024}
%
\begin{eqnarray}
\label{eq6}
&\Theta(k,t)&=\;\; \tan ^{-1}\left( \frac{-|b_{k}|^{2}\sin(2\epsilon_{k}^{f} t)}{|a_{k}|^{2}+|b_{k}|^{2}\cos(2\epsilon_{k}^{f} t)}\right), \\
\label{eq7}
&\Theta^{D}(k,t)&=\;\; -2|b_{k}|^{2}\epsilon_{k}^{f} \;,
\end{eqnarray}
%
so that
%
\begin{equation}
\label{eq8}
\Theta^{G}(k,t)= \tan^{-1}\left( \frac{-|b_{k}|^{2}\sin(2\epsilon_{k}^{f} t)}{|a_{k}|^{2}+|b_{k}|^{2}\cos(2\epsilon_{k}^{f} t)}\right)+2|b_{k}|^{2}\epsilon_{k}^{f}t \;. 
\end{equation}
%
In the following section, we revisit the phase diagram and exact solution of the one-dimensional transverse field Ising model with three spin interacting 
and also the noiseless DQPT.


\section{MODEL AND EXACT SOLUTION }\label{section3}

The Hamiltonian system under investigation arises from hybridisation between quantum statistical mechanics with quantum computation. 
A reference system is established using cold atoms in a triangular optical lattice \cite{Becker2010}. With an appropriate selection of parameters, 
this system can be modeled as a spin system exhibiting a specific ring-exchange interaction within the triangular lattice, which can subsequently 
be transformed into a "zig-zag chain". 
This setup creates a physical platform for a one-way route to quantum computation, where the algorithm involves specific measurements 
aimed at reconstructing the high degree of entanglement characteristic of the cluster state \cite{Briegel2001}. Interestingly, in addition 
to the three-spin ring-exchange interaction, various two-spin interactions can emerge in the system. Therefore, the cluster interaction competes 
with the exchange interaction by tuning a control parameter \cite{Pachos2004,Kopp2005,Montes2012,Niu2012,Skrovseth2009,Doherty2009,Son2011}.

The Hamiltonian of linear time dependent transverse field Ising model with cluster interaction \cite{Kopp2005,Niu2012} is given as
%
\begin{eqnarray}
\label{H1}
H\left( t \right) =  &-&{J}\sum_{j=1}^{N} \sigma _{j - 1}^z\sigma _j^z- h\left( t \right)\sum\limits_{j = 1}^N {\sigma _j^x}\\
\nonumber
&-&{J_3}\sum_{j=1}^{N}{\sigma _j^x\sigma _{j - 1}^z\sigma _{j + 1}^z},
\end{eqnarray}
%
where $\sigma^{\alpha}_{j} \;\;( \alpha= x, z)$ represents the Pauli matrices that act on site $j$  for
a chain of length $N$ with periodic boundary conditions (PBC) ($\sigma^{\alpha}_{N+1}=\sigma^{\alpha}_{1}$), $J$ denotes 
the strength of the nearest neighbor ferromagnetic interaction, $J_3$ indicates the strength of the three-spin interaction, and the transverse field 
$h(t) = h_{f} +vt$ ramping up from the initial value $h_{i}<0$ at time $t = t_{i}<0$ to the final values $h_{f}$ at $t_f \rightarrow 0^{-}$, 
with sweep velocity $v$ (Fig. \ref{fig1}). 
For $J_{3}=0$, the model reduces to the well-known transverse field Ising model.
%
\begin{figure}[t]
\centering{\includegraphics[width=\linewidth]{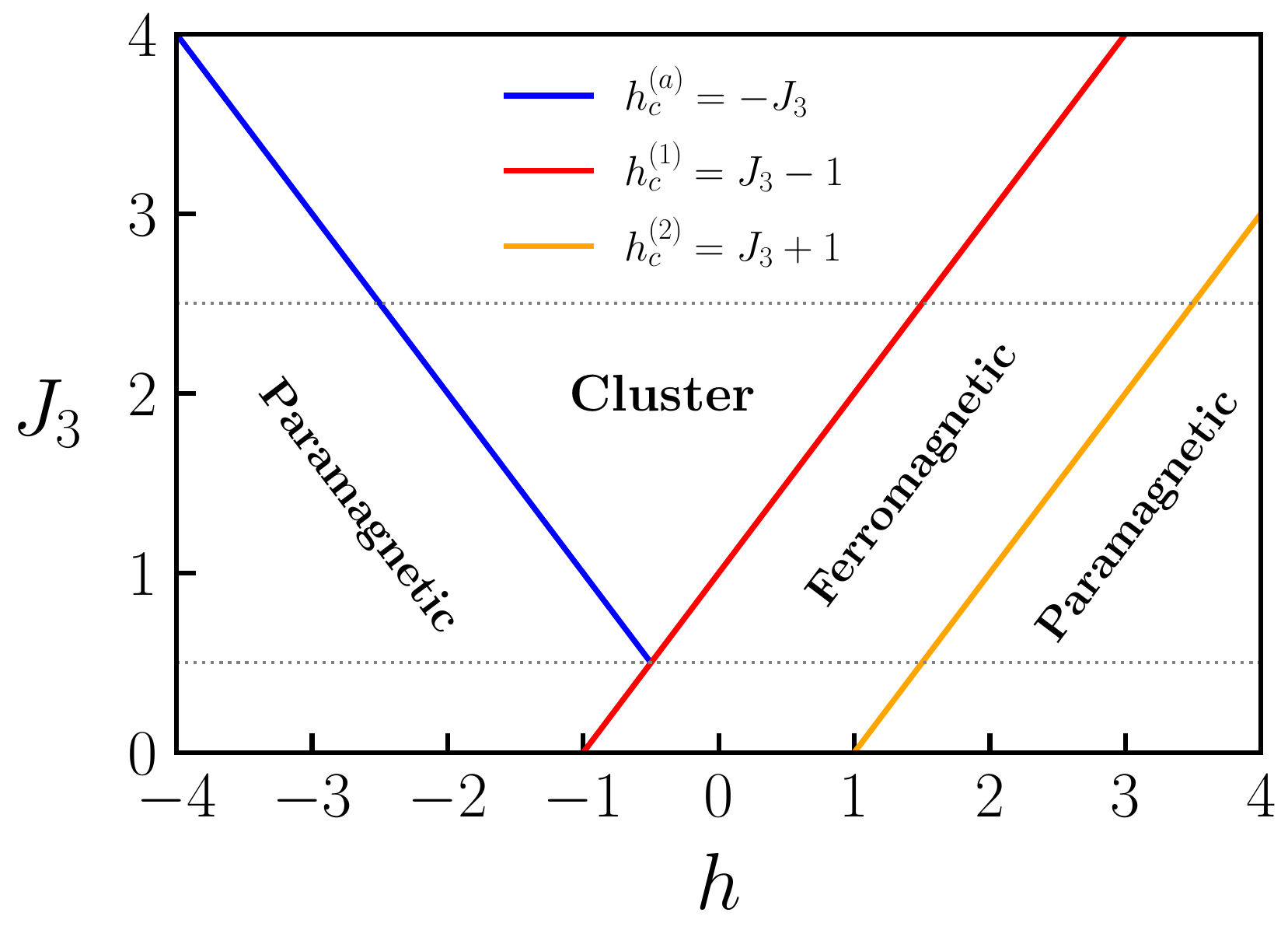}}
\caption{ (Color online) The equilibrium phase diagram of the Ising model with cluster interaction. 
The blue line indicates the anisotropy transition which takes place between the paramagnetic and cluster phases.
The red and yellow lines, represent the Ising like transition points.}
\label{phase-space-1}
\end{figure} 
%
By performing the Jordan-Wigner transformation 
%
\begin{eqnarray}
\nonumber
\sigma _j^z =  - \prod\limits_{j = 1}^{j - 1} {\sigma _j^x\left( {c_j^\dag  + {c_j}} \right)},~~~\sigma _j^x &=& 1 - 2c_j^\dag {c_j},
\end{eqnarray}
%
and applying the Fourier transformation ${c_j} =( e^{-i\pi/4}/{\sqrt N })\sum_k {{\exp{( - ikj)}}} {c_k}$, the Hamiltonian of Eq. \eqref{H1} 
can be expressed as the sum of $N/2$ non-interacting terms
%
\begin{eqnarray}
\label{H2}
H(t)=\sum_{k>0} H_{k}(t),
\end{eqnarray}
with
%
\small{
\begin{eqnarray}
\nonumber
H_{k}( t ) = {A\left( {k,t} \right)\left( {c_k^\dag {c_k} - c_{ - k} {c_{ - k}^\dag}} \right)}
+ B\left( k \right)\left( {{c_{ - k}}{c_k} + c_k^\dag c_{ - k}^\dag } \right),
\end{eqnarray}
}
%
where $A\left( {k,t} \right) = 2\left( {h\left( t \right) - {J}\cos k - {J_3}\cos 2k} \right)$ and $B\left( k \right) = 2\left( {{J}\sin k + {J_3}\sin 2k} \right)$, and the summation over k is limited to positive $k$ values in the form $k=(2n-1)\pi/N$ with $n=1,...,N/2$. 	 
The Bloch single particle Hamiltonian $H_{k}(t)$ can be  expressed  as:
%
\begin{eqnarray}
\label{Hk1}
{H_k}\left( t \right) = \left( {\begin{array}{*{20}{c}}
{A\left( {k,t} \right)}&{B\left( k \right)}\\
{ B\left( k \right)}&{ - A\left( {k,t} \right)} 
\end{array}} \right),
\end{eqnarray}
%
with the instantaneous eigenstates and eigenvalues 
%
\begin{eqnarray}
\label{eq12}
\varepsilon^{\pm}_{k}(t) &=& \pm \varepsilon_k(t)=\pm \sqrt{A^{2}(k,t)+B^{2}(k)},\\ [0.5em]
\nonumber
|\alpha_{k}(t)\rangle &=& \sin(\theta_{k}(t))|\up\rangle-\cos(\theta_{k}(t))|\down\rangle,\\[0.5em]
\nonumber
|\beta_{k}(t)\rangle &=& \cos(\theta_{k}(t))|\up\rangle+\sin(\theta_{k}(t))|\down\rangle,
\end{eqnarray}
%
where
%
\begin{eqnarray}
\nonumber
\sin(\theta_{k}(t))=\sqrt{\frac{1}{2}(1-\frac{A(k,t)}{\varepsilon_{k}(t)})},\\[0.5em]
\nonumber
\cos(\theta_{k}(t))=\sqrt{\frac{1}{2}(1+\frac{A(k,t)}{\varepsilon_{k}(t)})}.
\end{eqnarray}
%
	
Without loss of generality we set $J=1$ as the energy scale and $J_3>0$. For the time independent case ($h(t) = h$), the equilibrium phase diagram of the model can be constructed by identifying the regions of quantum criticality where the system becomes gapless in the thermodynamic limit ($N \to \infty$). 
It can be shown that the gap of spectrum vanishes at $h^{(1)}_c=J_3-1$ and $h^{(2)}_c=J_3+1$, with ordering wave vectors $k=\pi$ and $k=0$, respectively.
These two lines correspond to the quantum phase transitions from a quantum paramagnetic phase to a ferromagnetically ordered phase with the associated exponents 
being the same as the transverse field Ising model \cite{Divakaran2007}. Moreover, there is an additional gap closing point at $h_c^{(a)}=-J_{3}$. This transition
belongs to the universality class of the anisotropic transition observed in the transverse $XY$ model dual to the Hamiltonian in Eq. (\ref{H1}) \cite{Bunder1999}.
The phase boundary is surrounded by the incommensurate phases on either side with wave vector given by 
%
\begin{eqnarray}
\label{AM}
k_a=\arccos{((J_{3}-h)/4hJ_{3})}.
\end{eqnarray}
%

Obviously, for $J_{3}\le 1/2$, the anisotropic phase transition can not occur. The equilibrium phase diagram of the model is depicted in Fig. \ref{phase-space-1}. 
We will exploit an intriguing feature of the model: the movability of the gap-closing mode in the Brillouin zone, controllable by tuning the cluster interaction,
to uncover the aspects of DQPTs, under linear time driven magnetic field.

The time-dependent Schrödinger equation of the Hamiltonian Eq. \eqref{Hk1} with the linear time dependent transverse field,
can be transformed into the Landau-Zener (LZ) problem (see Appendix \ref{a}) which is exactly solvable \cite{Vitanov1999, Vitanov1996}. 
If the system prepared initially in its ground state at $ h_{i}= -10$ ($ t_{i} \to -\infty$), the probability of the $k$:th mode being in the excited 
state at a finite time $t$ is determined by the non-adiabatic transition probability \cite{Vitanov1999} (see Appendix \ref{a}).
 	
%
\begin{figure*}[]
\begin{minipage}{\linewidth}
\centerline{\includegraphics[width=0.33\linewidth]{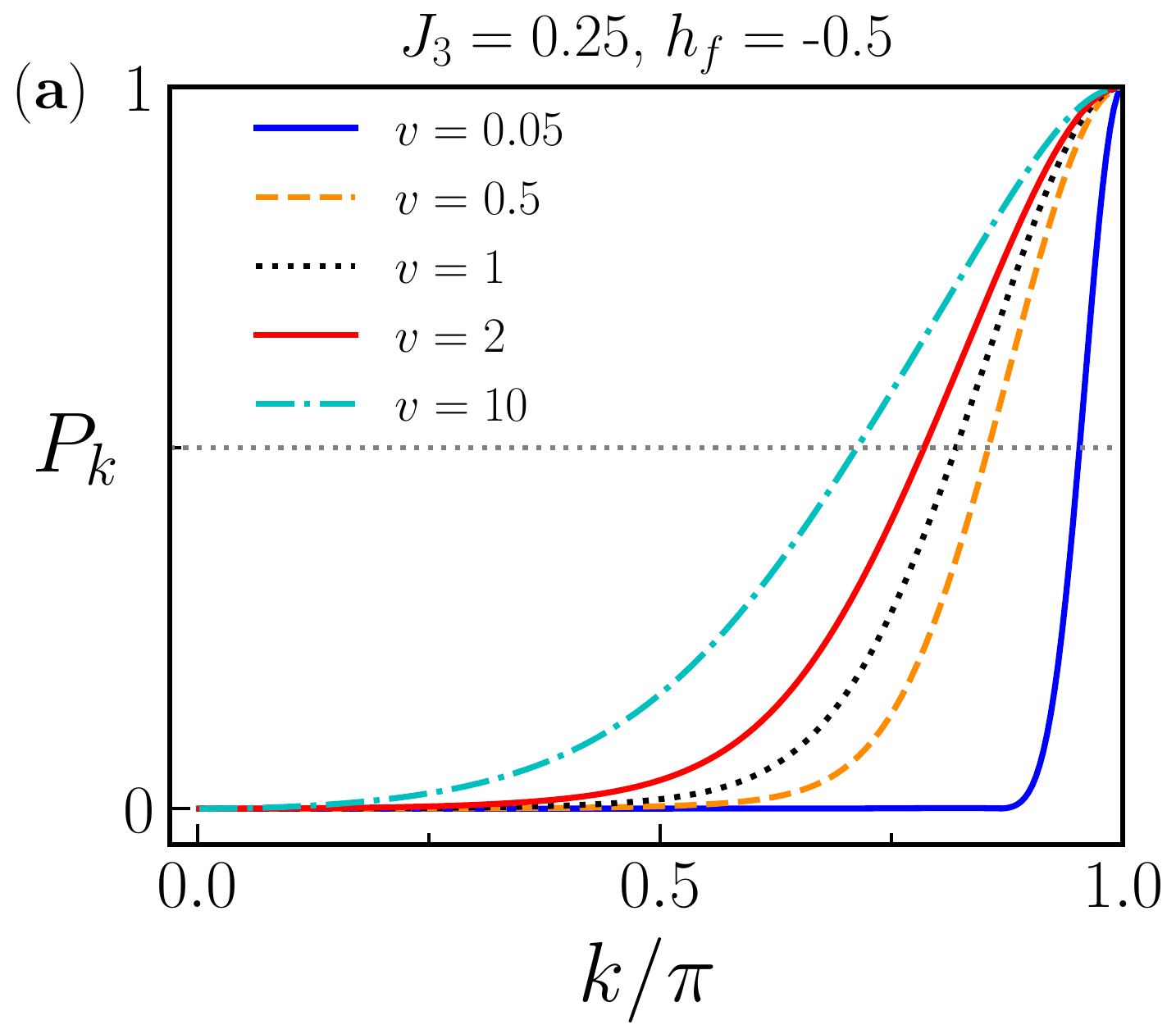}
\includegraphics[width=0.33\linewidth]{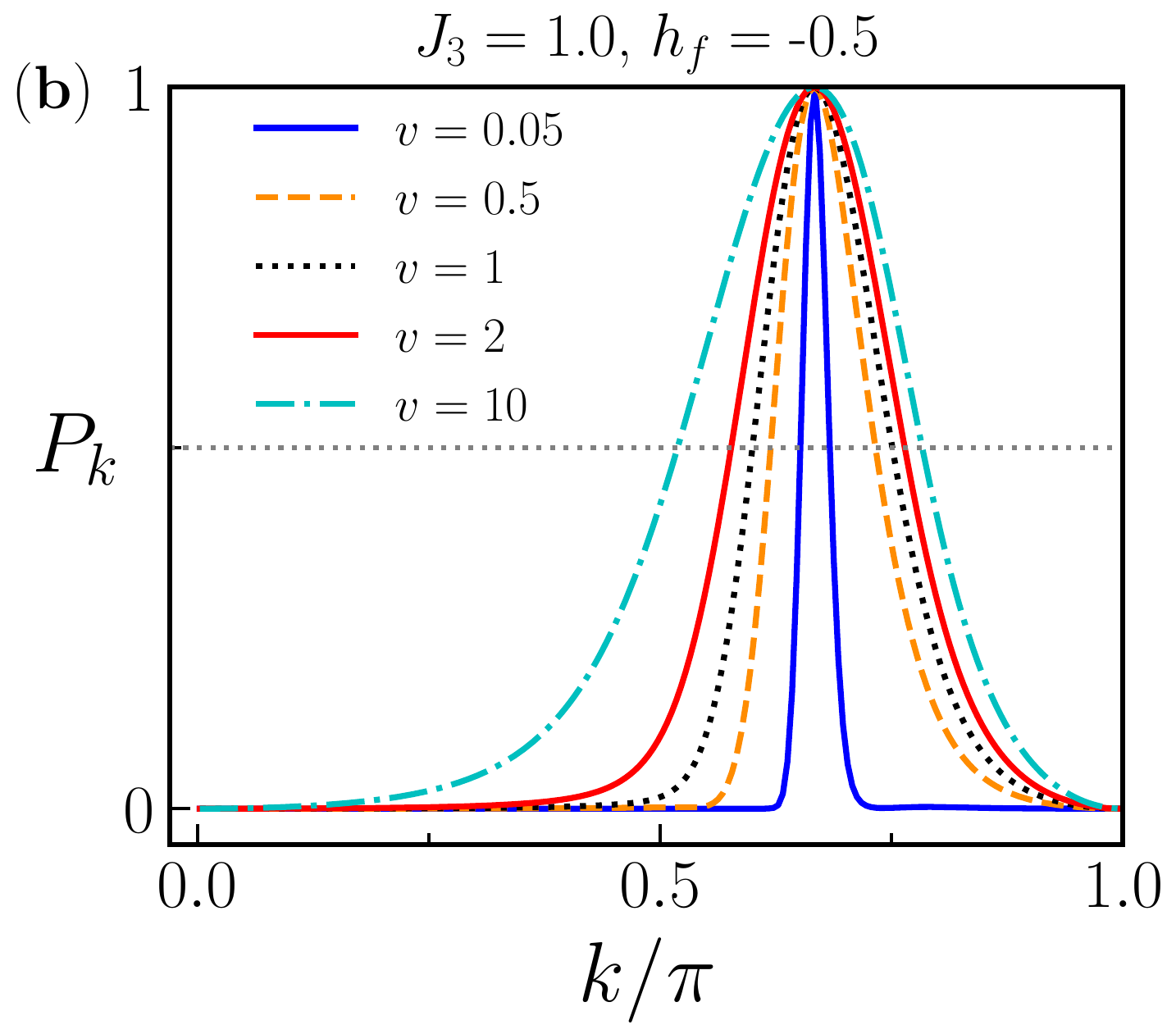}
\includegraphics[width=0.33\linewidth]{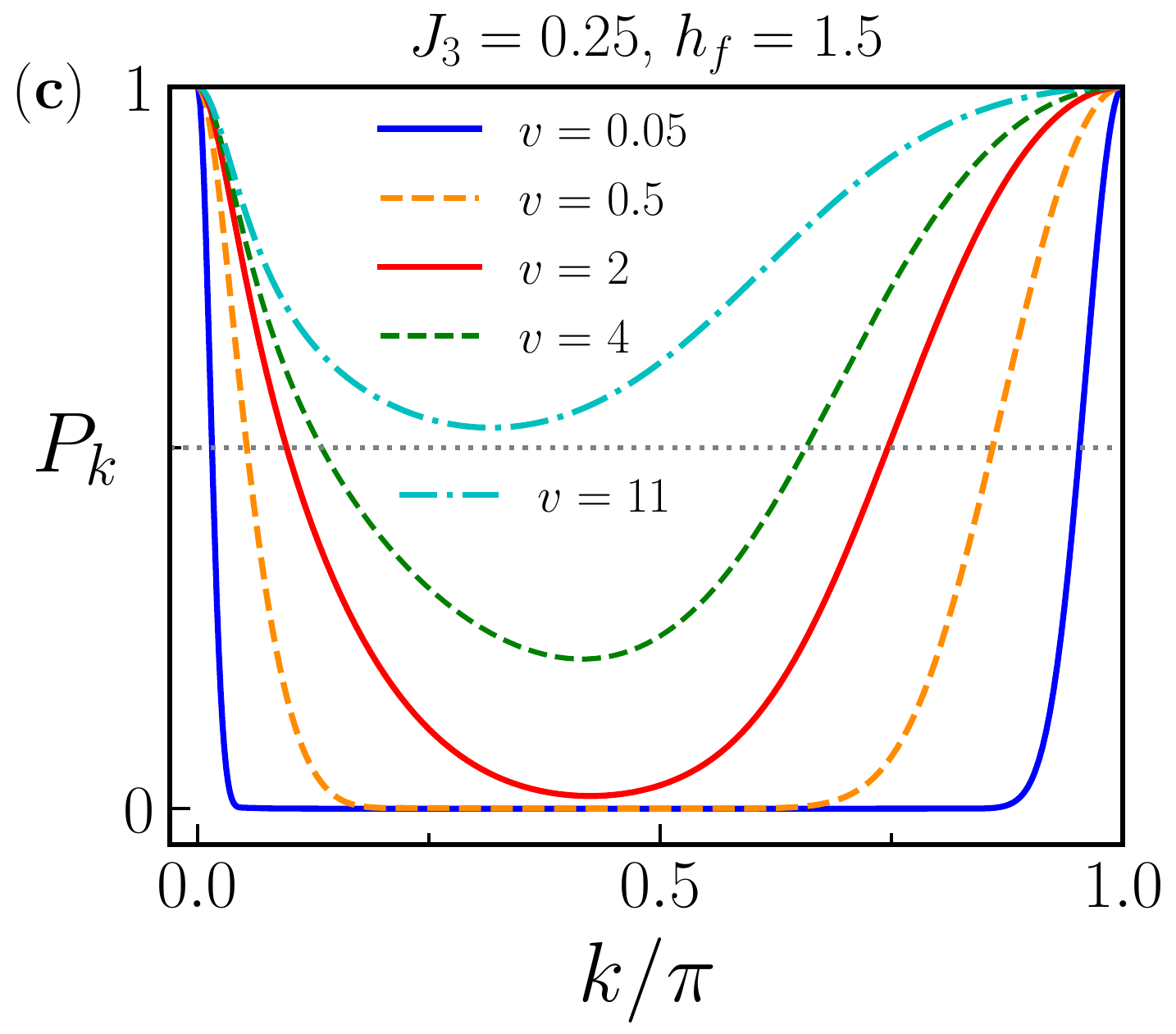}}
\centering
\end{minipage}
\begin{minipage}{\linewidth}
\centerline{\includegraphics[width=0.33\linewidth]{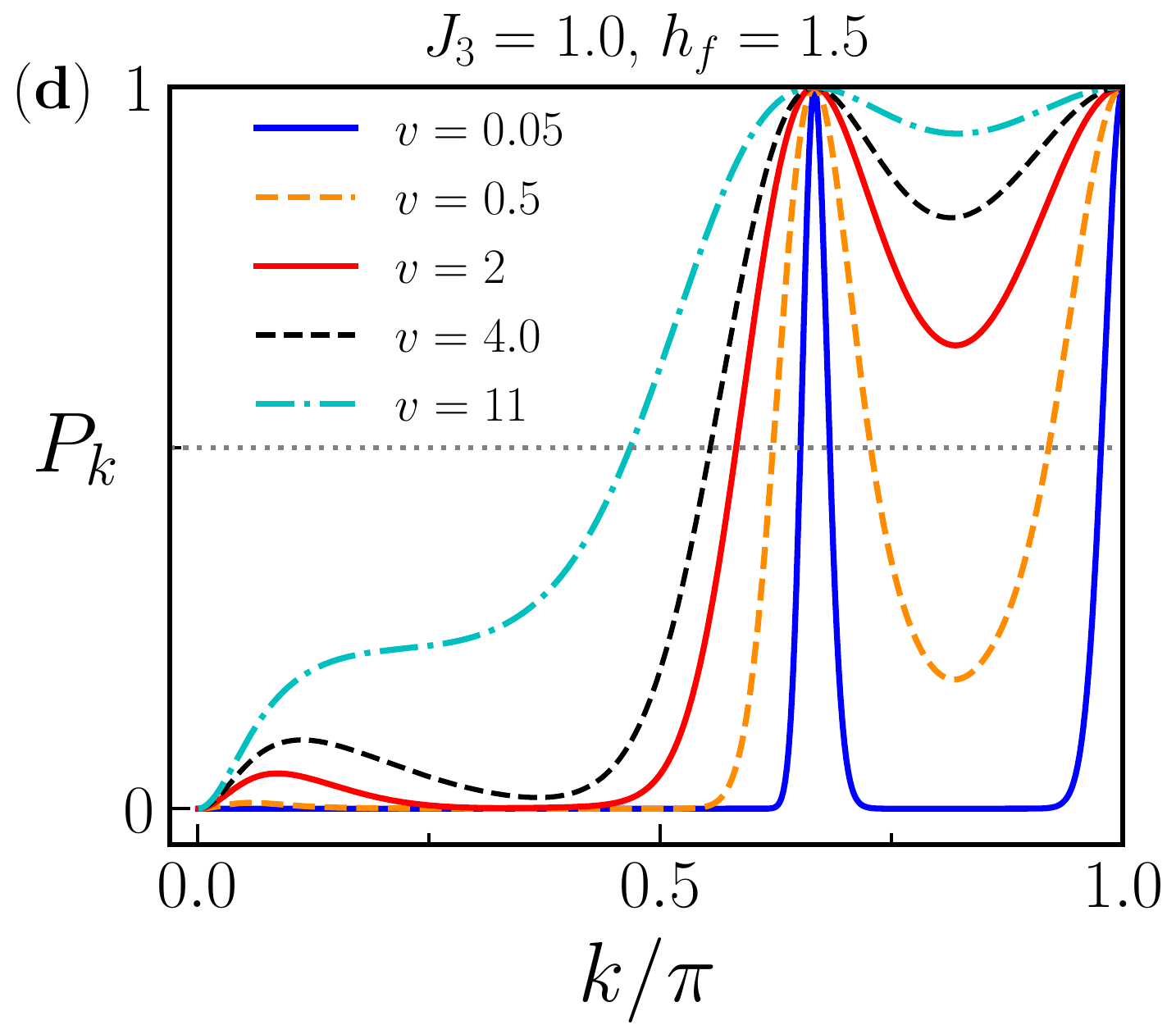}
\includegraphics[width=0.33\linewidth]{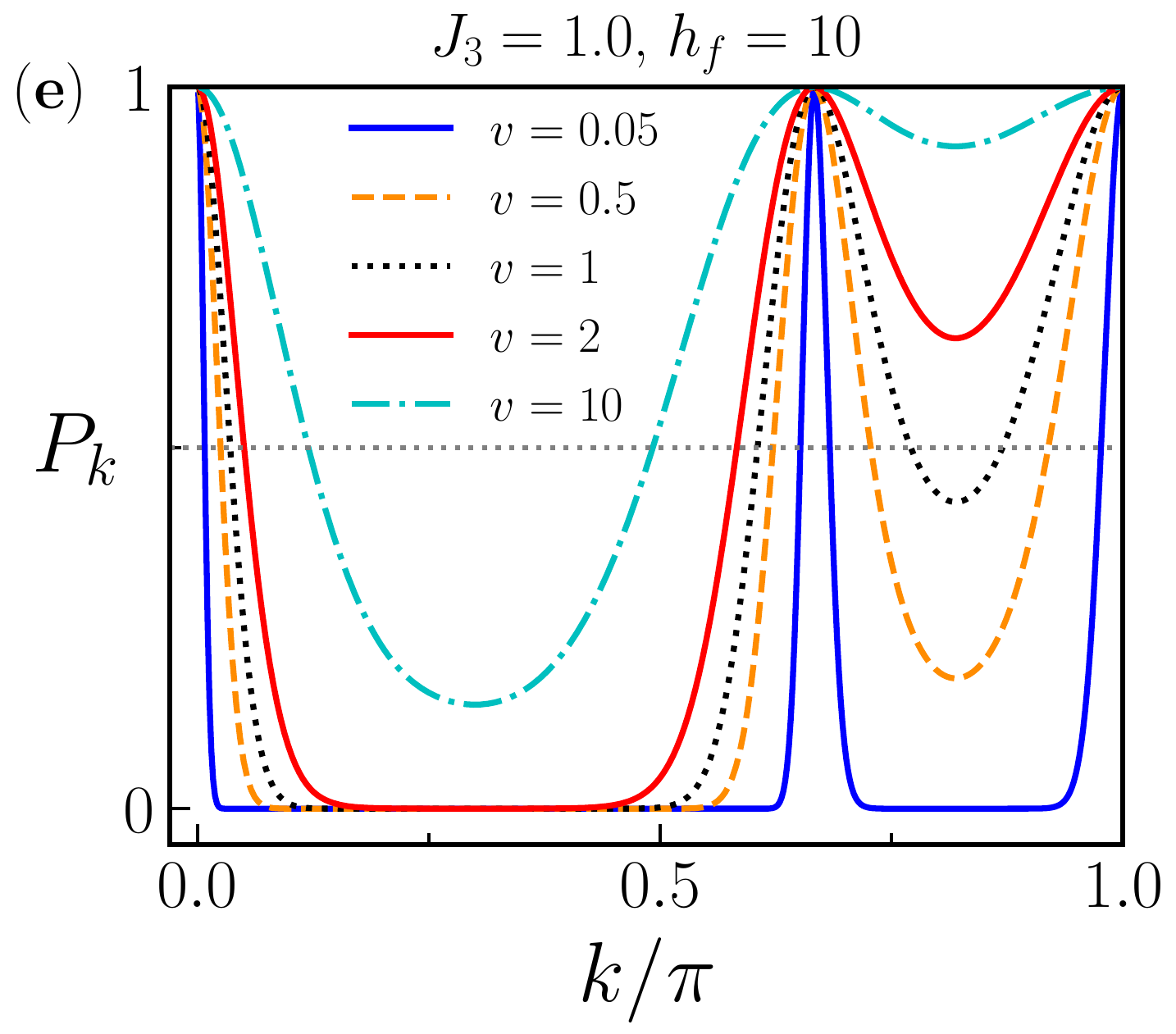}
\includegraphics[width=0.33\linewidth]{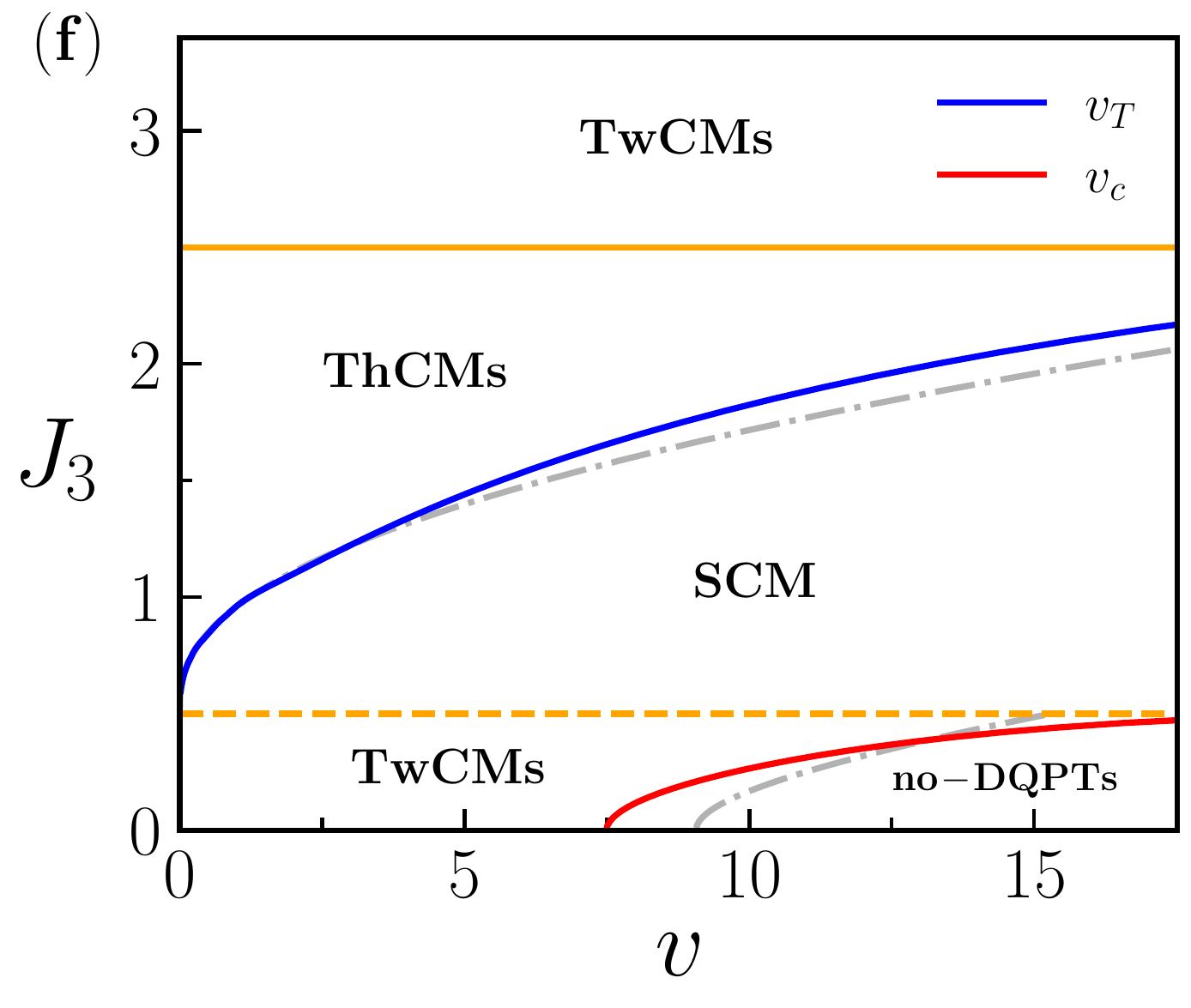}}
\centering
\end{minipage}
\caption{(Color plot) The transition probability $p_{k}$ 
following the quench from the initial value of transverse field $h_{i}=-10$ 
to the various values of quench field end $h_{f}$ for various sweep velocities. 
(a) for $J_{3}=1/4$ and $h_{f}=-0.5$ where the ramped quench 
crosses a single Ising like quantum critical point $h^{(1)}_{c}=-3/4$, (b) for 
$J_{3}=1$ and $h_{f}=-0.5$ where the ramped quench crosses single 
quantum critical points  $h^{(a)}_{c}=-1$, (c) for 
$J_{3}=1/4$ and $h_{f}=1.5$ where the quench crosses two Ising like 
quantum critical points $h^{(1)}_{c}=-3/4$, and $h^{(1)}_{c}=5/4$ (d) for $J_{3}=1$ and 
$h_{f}=1.5$ where the ramped quench crosses two quantum critical 
points  $h^{(a)}_{c}=-1$ and $h^{(1)}_{c}=0$,  (e) for $J_{3}=1$ and $h_{f}=10$, 
where the ramped quench crosses three quantum critical points  
$h^{(a)}_{c}=-1$, $h^{(1)}_{c}=0$, and $h^{(2)}_{c}=2$. 
(f) The dynamical phase diagram of the model in the $(J_{3}; v)$ plane in 
the absence of the noise for a quench from $h_i=-10$ to $h_{f}=1.5$. The diagram is divided into 
three regions. The horizontal yellow dashed-dotted line represents $J_3=1/2$ and $J_3=5/2$.
The blue solid line separates two regions where the system shows SCM and ThCMs. 
The red solid line shows the boundary between regions characterized by TwCMs and those with no-DQPTs. 
The blurred gray dashed-dotted curves represent the boundaries obtained from LZ transition formula.}
\label{Fig3}
\end{figure*}
%

%
\begin{eqnarray}
\label{eqTP}
p_{k}(\tau)&=& e^{\pi \gamma^{2}/2v} \Big|{\cal D}_{i\gamma^{2}/v} \Big( \sqrt{v} e^{3i\pi/4}\tau \Big)\cos(\theta_{k}(\tau))\\
\nonumber
&-&\frac{\gamma}{\sqrt{v}}e^{-i\pi/4} {\cal D}_{-1+i\gamma^{2}/v} \Big( \sqrt{v} e^{3i\pi/4}\tau \Big)\sin(\theta_{k}(\tau) \Big|^2\nonumber 
\end{eqnarray}
%
where $\tau=2(vt-\cos k-\lambda_{2}\cos 2k)/v$ , and $\gamma=B_{k}/2$, and ${\cal D}_{\nu}(z)$ is the parabolic cylinder function \cite{erdelyi1953,Abramowitz1988}. 
Furthermore, as $t_{f} \to +\infty$ ($h_{f} \gg h_c$), the probability of excitations $p_{k}$, is represented by the LZ transition probability,
%
\begin{eqnarray}
\label{pk-inf}
p_{LZ}=e^{-2\pi \gamma^2/v}.
\end{eqnarray}
In the following section, we will study the dynamical phase diagram of the model for the passage of the noiseless transverse field through the critical points.

 
\section{NOISELESS NUMERICAL RESULTS}\label{section4} 
	 
The results of our numerical simulations, conducted using an analytical approach, are presented in this section to analyze the dynamics 
of the model through the concept of DQPTs. For this aim, we consider three cases of the noiseless ramp protocol through the single, 
two and three critical points for quenches starting at $h_{i}=-10$.

		 	 
\subsubsection{Quench across a single critical point ($h_f=-1/2$)}\label{QASCP} 
When time driven magnetic field crosses a single critical point, i.e, $h_f=-1/2$, we consider two cases $J_{3}<1/2$ and $J_{3}>1/2$, respectively. 

\begin{description}
\item[(i)] 
In the case that $J_{3}<1/2$, the ramp field crosses the Ising type single critical point $h_{c}^{(1)}$, at $k_{c}=\pi$, and 
the probability of excitation $p_{k}$, depends on the value of $k$. As expected, when the system crosses the critical point, 
it undergoes non-adiabatic evolution due to the gap closing and yields maximum transition probability at the gap closing mode $p_{k =\pi}=1$.
Nevertheless, far away from the gap closing mode, the system evolves adiabatically due to the non-zero energy gap and leads to small transition probability
($p_{k=0} \to 0$). 
Given these two cases and continuity of the transition probability as a function of allowed mode $k$ in the thermodynamic limit, 
implies that there exist a critical mode $k^*$ at which $ p_{k^*} = 1/2 $ and consequently DQPTs always occur for a quench crosses a single critical point. 
The numerical simulation of transition probability has been plotted in Fig. \ref{Fig3}(a) versus $k$ for $h_{f} = - 1/2$ and $J_{3}=1/4$, for various sweep velocities 
as the ramp filed passes through the single critical point $h_{c}=-3/4$. As seen, for a quench that crossing a single critical point, 
there is always a critical momenta $k^*$ and consequently those of $t^{*}_n$, given by Eq. \eqref{t*}.

\item[(ii)] 
In the case that $J_{3}>1/2$, the ramp transverse filed crosses the single anisotropic transition point $h_{c}^{(a)}$, where the band gap closes 
inside the Brillouin zone at $k_{a}$ given by Eq. \eqref{AM}. At this critical point, the system undergoes a phase transition from the paramagnetic 
phase to the cluster phase. 
We expect that, the system experiences non-adiabatic transition at the gap closing mode, i.e., $p_{k_a} = 1$, and evolves adiabatically away from the gap closing mode
($p_{k=0,\pi} \to 0$). Since the gap closing occurs at interval $0<k_a<\pi$, expected that two critical modes emerge in the system at which $ p_{k^*} = 1/2 $. 

Fig. \ref{Fig3}(b) displays the transition probability versus $k$ for $J_3=1$ and $h_{f} = -1/2$, as 
the ramp field passages across the single critical point $h_{c}=-1$ where the band gap closes at $k=2\pi/3$. As seen, there are two critical modes $k^{*}_{\alpha}$ and $k^{*}_{\beta}$ at which $p_{k^{*}_{\alpha}}=p_{k^{*}_{\beta}} = 1/2$. In other words, if the quench crosses a single critical point at which the energy gap 
closes inside the Brillouin zone, the system reveals two critical modes at which DQPTs happen. While, for a quench that crossing 
a single critical point where the gap closing occurs at high-symmetry points in the Brillouin zone ($k=0, \pi$), the system includes only a single critical mode.  
 
\end{description}

%
\begin{figure}[t!]
\centering
\hspace{-1cm}\includegraphics[width=0.9\linewidth]{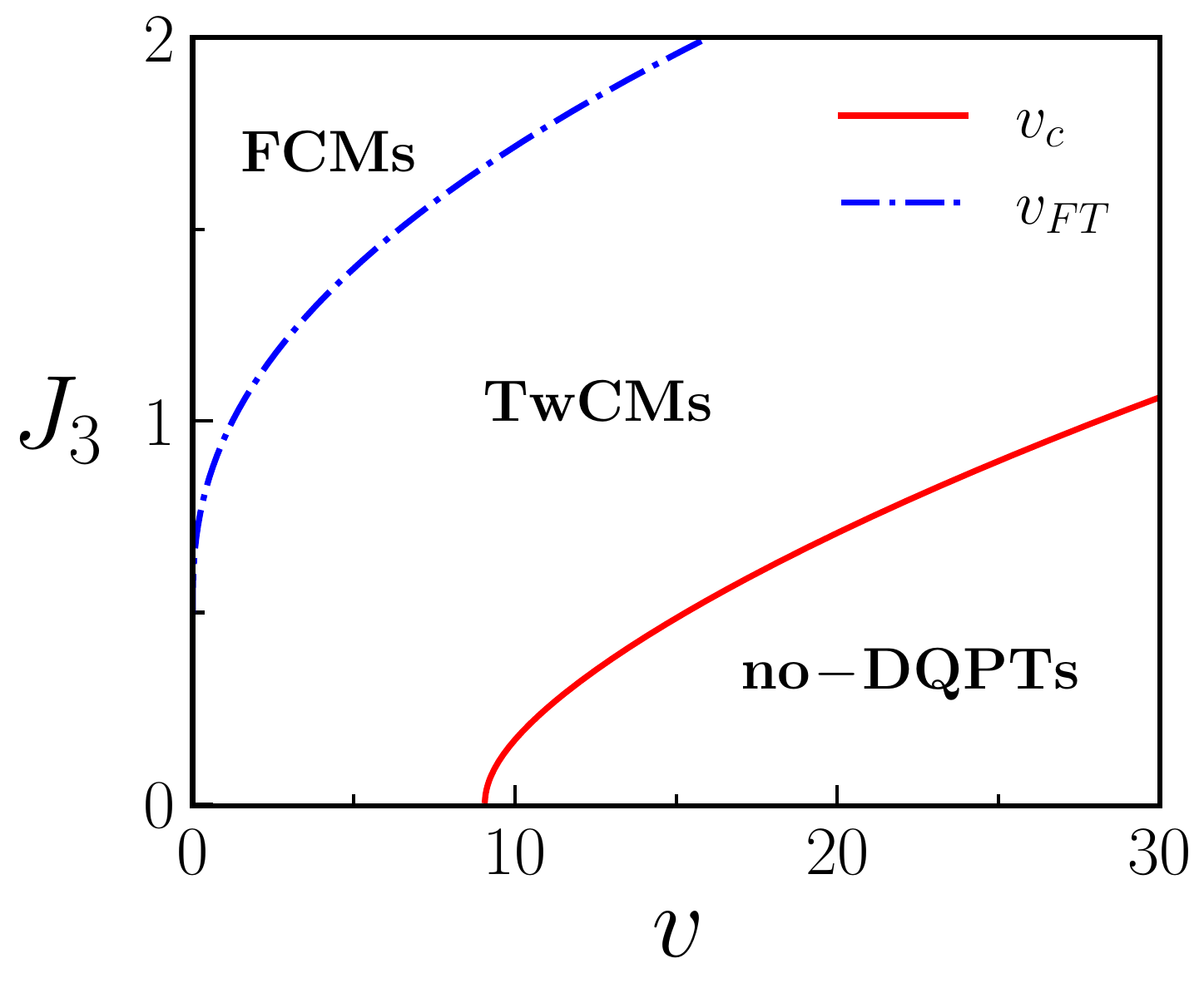}
\caption{(Color online) The dynamical phase diagram of the model in the $(J_{3}; v)$ 
plane for a quench from $h_i=-10$ to $h_f=10$ in the absence of the noise.}
\label{Fig4}
\end{figure} 
%
 
 
\subsubsection{Quench across the two critical points ($h_f=3/2)$}\label{QASTCP}  
For a quench crosses two critical points, i.e, $h_f=3/2$, we have considered once again two cases $J_{3}<1/2$ and $1/2<J_{3}<5/2$, respectively. 
It should be mentioned that, for $h_f=3/2$ and $J_3>5/2$, the quench transverse field crosses only a single anisotropic transition point and the
dynamics of the system is similar to that of discussed in previous section (\ref{QASCP}-(ii)).

\begin{description}
\item[(i)] 
In case of $J_{3} < 1/2$, the quench field crosses both Ising type critical points $h_{c}^{(1)}$ and $h_{c}^{(2)}$. In such a case, the field is swept from one equilibrium paramagnetic ground state to another. It should be mention that, If the quench is sudden it is not expected to observe the DQPTs \cite{Vajna2014,Sharma2016,Dora2013}. 
As anticipated, when the field crosses both critical points, the nonadiabatic evolution of the system at gap closing modes $k = 0, \pi$, leads to the maximum transition probability, i.e., $p_{k=0}=p_{k=\pi} = 1$. However, the minimum transition probability occurs approximately at the maximum energy gap mode. Since the maximum value of the transition probability $p_{k=0,\pi} = 1$, is greater than $1/2$, the emergence of DQPTs requires that the minimum transition probability becomes less than $1/2$.
Consequently, making the ramp sufficiently slow ensures that the minimum excitation probability becomes smaller than $1/2$.
In that case, there are two critical modes $k^{*}$ where $p_{k^*}=1/2 $, and consequently DQPTs appear.

In Fig. \ref{Fig3}(c) the transition probability has been plotted versus $k$ for $J_3=1/4$ for a quench that crossing two Ising like transition points at
$h_{c}^{(1)}=-3/4$ and $h_{c}^{(2)}=5/4$. As seen, $p_{k=0,\pi}=1$ and the minimum of $p_k$ away from the critical modes is less than $1/2$ for the small sweep velocity.
Therefore, there are two critical modes $k^{\ast}_{\gamma}$ and $k^{\ast}_{\delta}$ at which $p_{k^{\ast}_{\gamma,\delta}}=1/2$, yielding DQPTs at the corresponding critical times $t^{\ast}_n=t^{\ast}_{n,\gamma}, t^{\ast}_{n,\delta}, n=0,1,\ldots$.
However, the minimum of $p_k$ is greater than $1/2$ for a sweep velocity larger than the critical sweep velocity $v>v_c(h_f,J_3)$, and consequently DQPTs are wiped out.

The critical sweep velocity above which DQPTs disappears, can be calculated analytically for $h_{f}\gg h_c^{(2)}$, using the LZ transition probability Eq. (\ref{pk-inf}). 
As anticipated, at the gap closing modes ($k=0,\pi$), the LZ transition probability is maximum $p_{k=0,\pi}=1$, and the minimum of $p_{LZ}$ occurs at $k_{m}=\arccos ((-J + \sqrt{J^{2}+32J_{3}^{2}})/8J_{3})$. Therefor, DQPTs appear if $p_{k_m} \leqslant 1/2$, which satisfied for $v\leqslant v_c$ with $v_{c} = 2\pi \gamma_{m}^{2} / \ln (2)$ where $\gamma_{m} = {{J}\sin k_{m} + {J_3}\sin 2k_{m}}$.
Thus, for a quench crosses both Ising like transition points, there exists a critical sweep velocity $v_c$ above which DQPTs are eliminated and DQPTs do not appear for a sudden quench \cite{Sharma2016,Zamani_2024,Jafari2024,Baghran2024}.

\item[(ii)] 
In case of $J_3>1/2$, the transverse field is swept from one equilibrium paramagnetic phase to the ferromagnetic phase. 
In such a case, the system crosses the anisotropic ($h_{c}^{(a)}$) and Ising like ($h_{c}^{(1)}$) transition points at $k_a$ and $k =\pi$, 
which yielding $p_{k_a}=p_{k=\pi} = 1$. However, we expect small transition probability for modes away from the gap closing modes. 
Since, the maximum value of transition probability $p_{k=0,\pi}=1$, is greater than $1/2$, and the modes away from the gap closing mode
evolves adiabatically, i.e., $p_{k\rightarrow 0}\rightarrow 0$, we can conclude that there should exist critical mode at which $p_{k^\ast}=1/2$
and consequently DQPTs occur.

In Fig. \ref{Fig3}(d) the transition probability has been shown versus $k$ for $h_{f} = 3/2$ and $J_3=1$, for different values of sweep velocity. 
As clear, the transition probability is maximum at the gap closing modes $p_{k=2\pi/3,\pi}$, and $p_k$ remains zero at $k=0$, for any values of sweep velocity. 
Therefor, in contrast to $J_3<1/2$ case, the critical mode always appears in the system and DQPTs take place even for a sudden quench case. 

The surprising result occurs for the modes between two gap closing modes ($2\pi/3<k<\pi$). As observed, while the transition 
probability takes it maximum value at gap closing modes, the minimum of transition probability becomes less than $1/2$ 
for slow ramping. Thus, the system can encompasses three critical modes for a sweep velocity below the sweep velocity which named triple sweep 
velocity $v_T$.

It is worth to mention that, although the LZ transition probability convinced for $h_f\gg h_c$, our numerical simulation shows that the transition probability for $2\pi/3<k<\pi$ can be approximately is described by the LZ transition probability. 

\end{description}

Comparing these two cases reveals that the modes confined between two gap closing modes, can be easily excited to the upper level for large sweep velocity. 
However, the modes which do not restricted between two modes with maximum transition probability $(p_k=1)$, prefer to stay at the lower level even for a sudden quench case. 
Although in both cases the quench field crosses two critical points, neither starting point nor ending point of the ramp field 
is confined between two critical point for $J_3<1/2$ case. While for $J_3>1/2$ case, ramp field end is restricted between two critical points. Therefore, we
come to the conclusion that, DQPTs are always present if starting or ending point of the quench field is restricted between two critical points, even for sudden quench case.
Otherwise, there is critical sweep velocity above which DQPTs are wiped out.   

%
\begin{figure*}[]
\begin{minipage}{\linewidth}
\centerline{\includegraphics[width=0.33\linewidth]{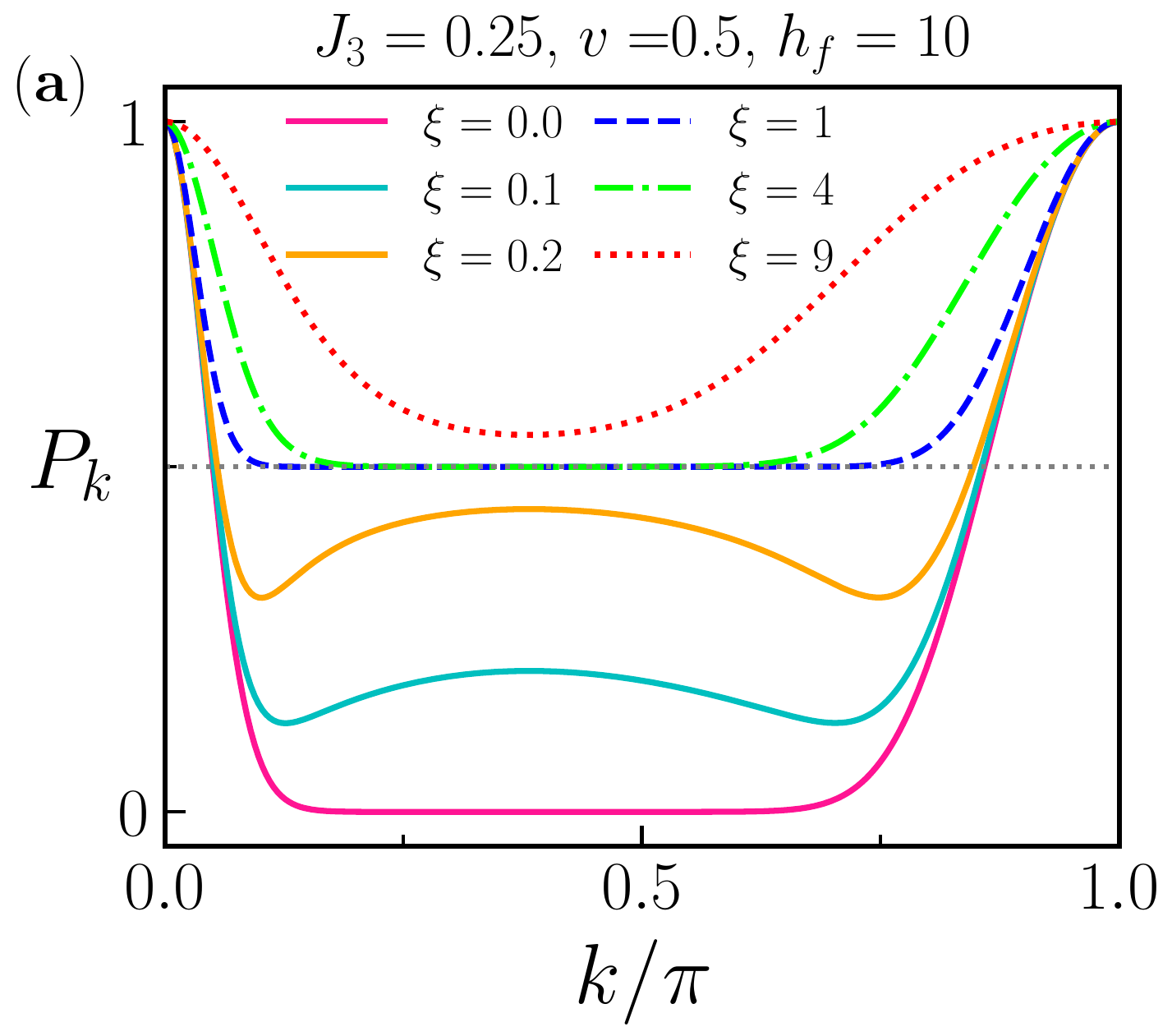}
\includegraphics[width=0.33\linewidth]{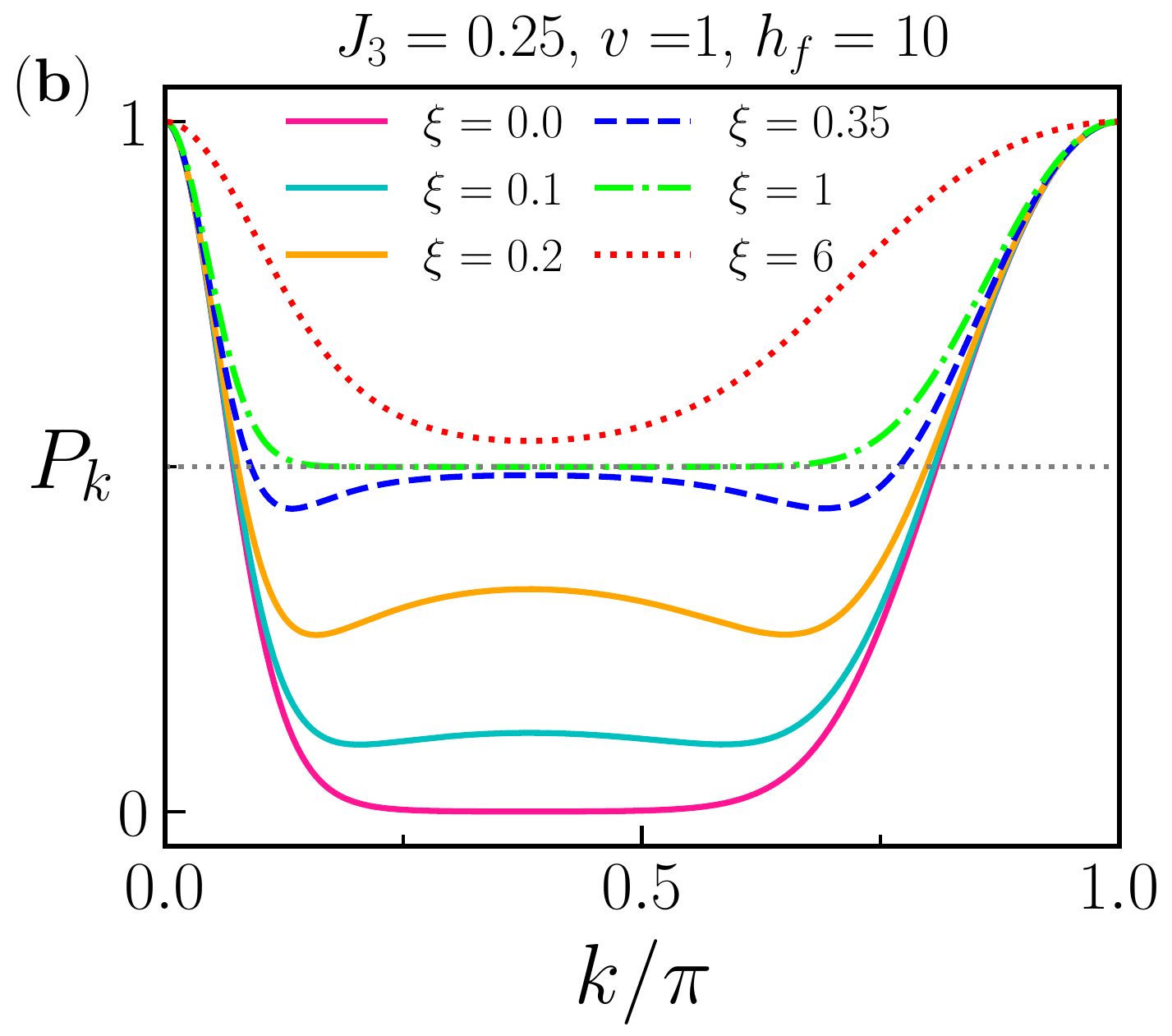}
\includegraphics[width=0.33\linewidth]{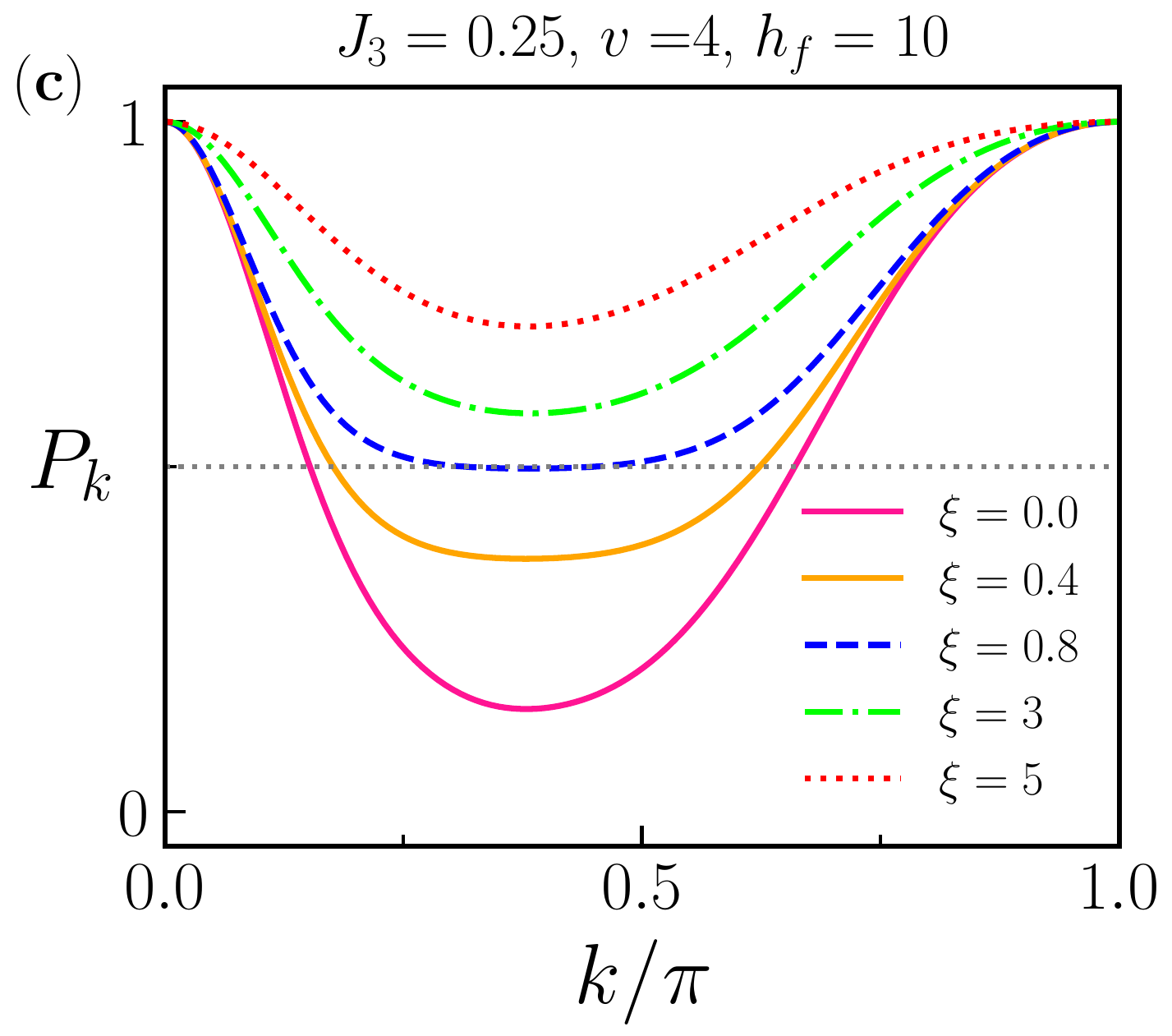}}
\centering
\end{minipage}
\begin{minipage}{\linewidth}
\centerline{\includegraphics[width=0.33\linewidth]{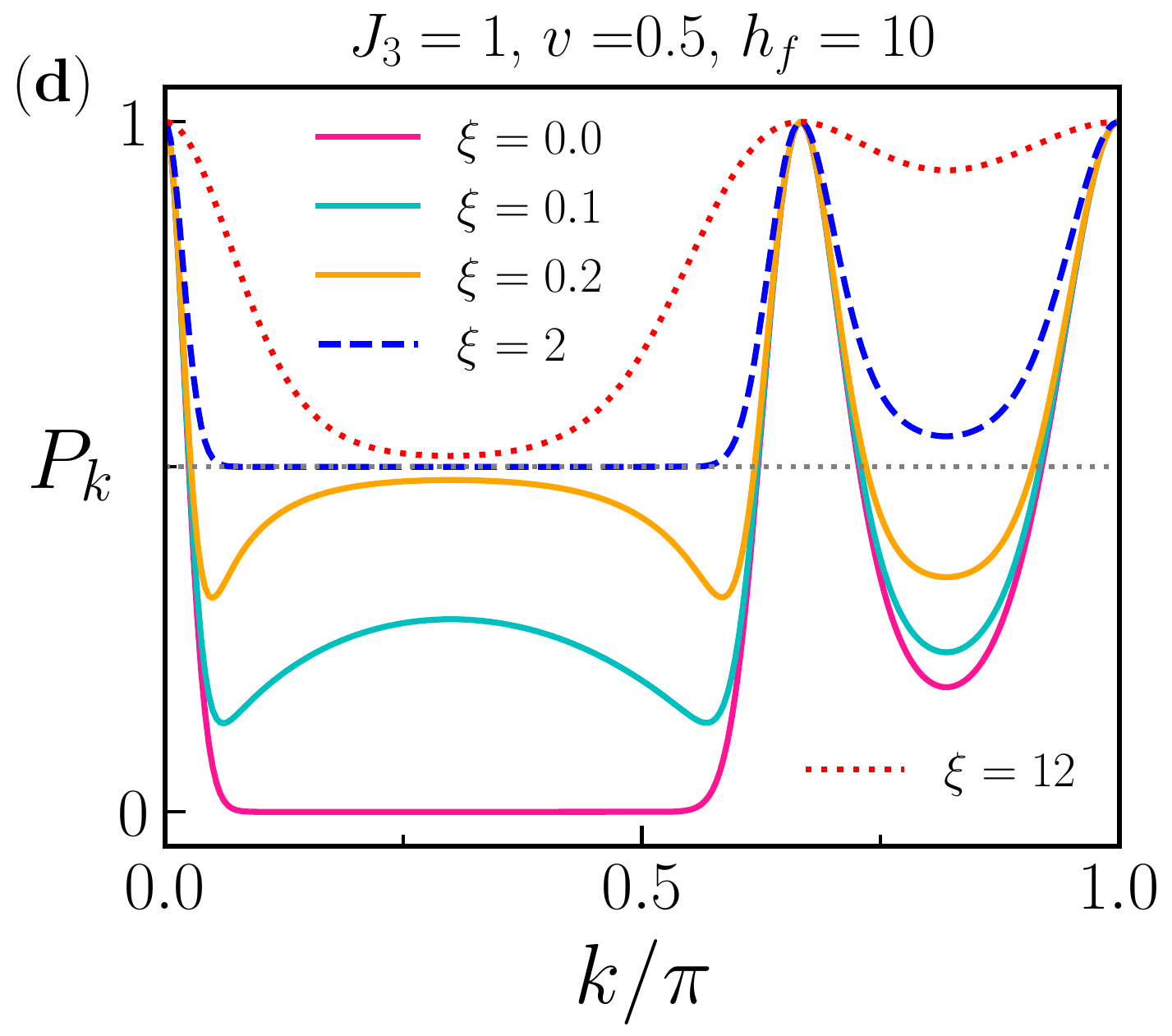}
\includegraphics[width=0.33\linewidth]{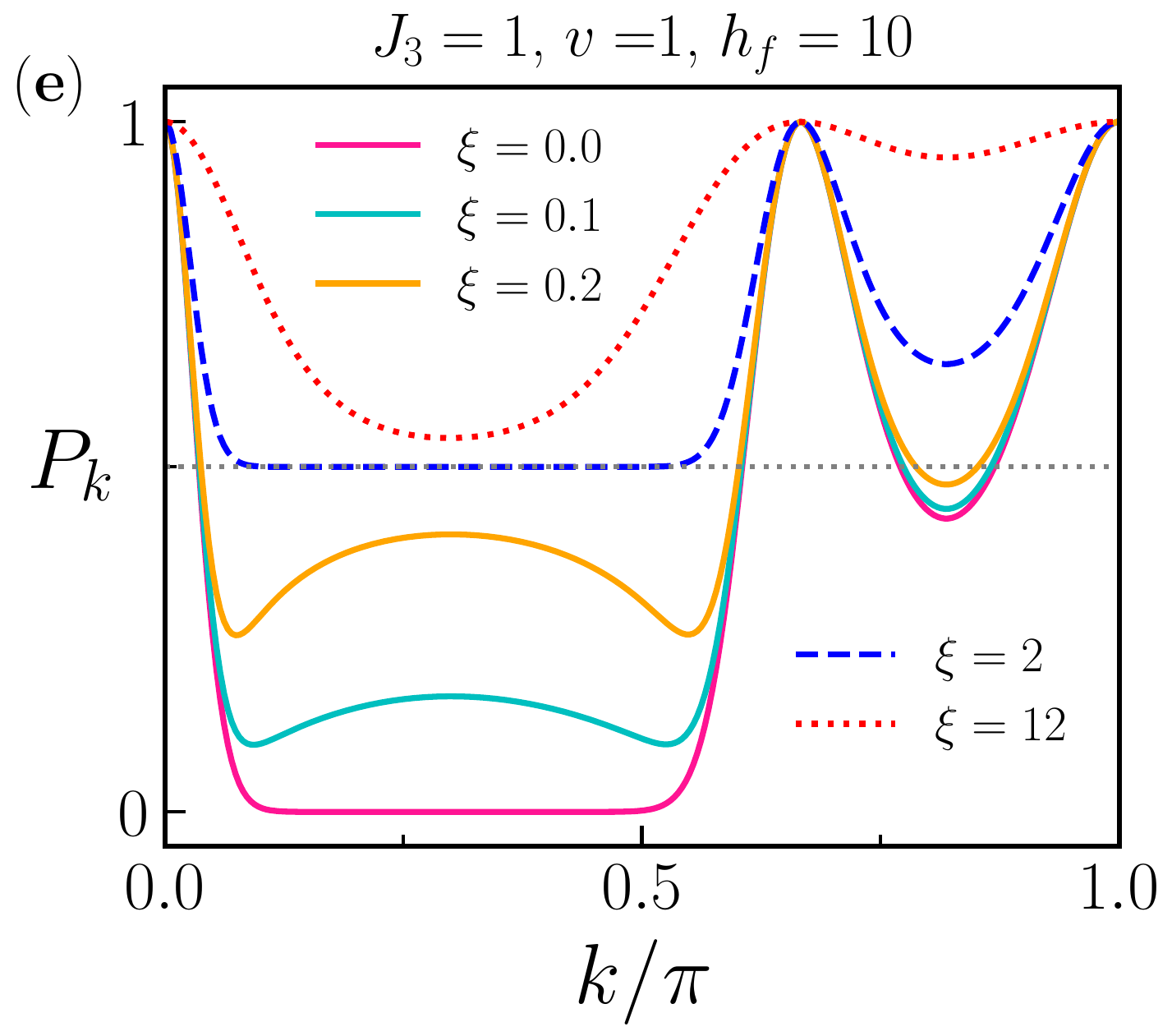}
\includegraphics[width=0.33\linewidth]{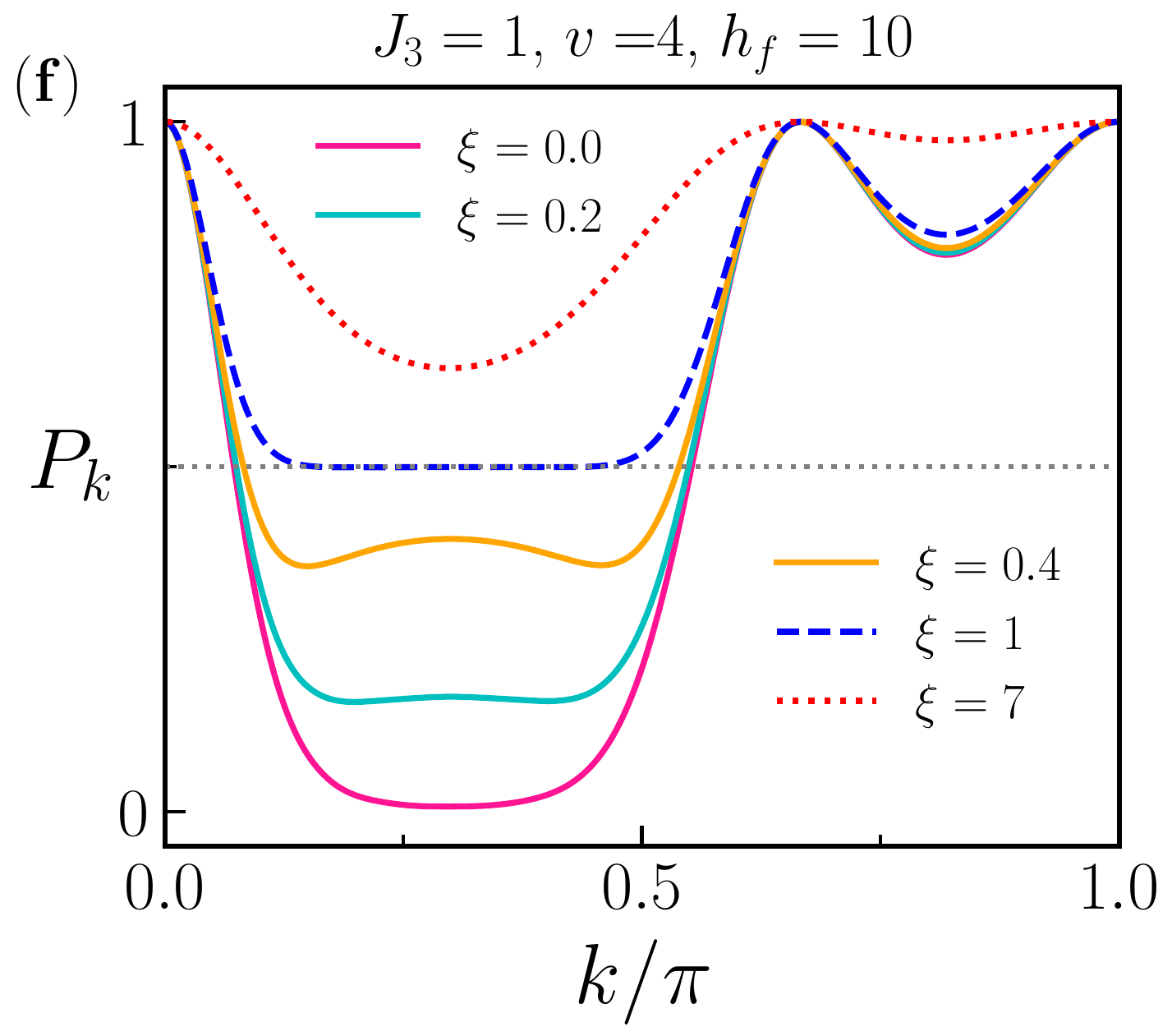}}
\centering
\end{minipage}
\caption{(Color plot)  The probability of excitations for a quench from $h_i=-10$
to $h_f=10$ for different values of noise strength for $J_{3}=1/4$: (a) $v = 0.5$, (b) $v = 1$, 
and (c) $v = 4$, as well as for $J_{3}=1$:  (d) $v = 0.5$, (e) $v = 1$, and (f) $v = 4$.}  
\label{Fig5}
\end{figure*}
%

	 	
\subsubsection{Quench across the three critical points ($h_f=10$)}
To perform a quench across three critical points, the quench filed end should be larger than $h_c^{(2)}$ and the cluster interaction $J_3>1/2$.
In such a case, the quench field is swept from one equilibrium paramagnetic phase to another one and passage through the critical points 
$h_c^{(a)}$, $h_c^{(1)}$ and $h_c^{(2)}$ where the band gap closes at $k=k_a$, $k=\pi$ and $k=0$. Thus, the transition probability is maximum
at $k=k_a$, $k=\pi$ and $k=0$ and $p_k$ discloses two minimum between three gap closing modes. Appearance of DQPTs is required that the minimums of $p_k$ 
becomes less than $1/2$. In this case, the system encompasses four critical modes and DQPTs is wiped out for a sweep velocity above the critical sweep velocity.
Moreover, we expect that the system shows transition from four critical modes case to two critical modes case for sweep velocity 
smaller than the critical sweep velocity. 
  
For $h_{f}\gg h_c^{(2)}$, the transition probability is given as LZ transition probability and the critical sweep velocity above which DQPTs disappear
can be obtained analytically. As mentioned, DQPTs appear if the condition $p_{LZ}^{min} \leqslant 1/2$ is satisfied.
It is straightforward to show that, two minimum of the transition probability occurs at $k=k^{\pm}_{m}=\arccos ((-J \pm \sqrt{J^{2}+32J_{3}^{2}}) / 8J_{3})$.
More analysis manifest that, the minimum at $k^{+}_{m}$ is the global minimum of transition probability while the minimum at $k^{-}_{m}$ is the local minimum.
Consequently, if both global and local minimums of $p_k$ are less than $1/2$ the system illustrates four critical modes which yielding DQPTs. 
Further, the transition from four critical modes case to two critical modes case occurs if the local minimum of $p_k$ is greater than $1/2$ while 
DQPTs are removed if the global minimum exceed $1/2$. 
  
Detailed analysis shows that, the condition $p_{k^{+}_m} \leqslant 1/2$ is satisfied if $v\leqslant v_c$ with $v_{c} = 2\pi (\gamma^{g}_{m})^{2} / \ln (2)$ where $\gamma^{g}_{m} = {{J}\sin k^{+}_{m} + {J_3}\sin 2k^{+}_{m}}$ which results DQPTs. In addition, the system transits from four critical modes case to two critical modes case for a sweep velocity $v>v_{FT}=2\pi (\gamma^{\ell}_{m})^{2} / \ln (2)$ where $\gamma^{\ell}_{m} = {{J}\sin k^{-}_{m} + {J_3}\sin 2k^{-}_{m}}$.

The transition probability has been shown in Fig. \ref{Fig3}(e) versus $k$, for $h_f=10$ and $J_3=1$. As seen, $p_k$ represents two minimum
at $k^{+}_m/\pi=0.299$ and $k^{-}_m/\pi=0.819$, and for sufficient slow sweep velocity system unveils four critical modes at which $p_{k^{\ast}}=1/2$. 
As the sweep velocity increases, the transition probability at $k^{\pm}_m$ enhances and $p_{k^{-}_m}$ exceeds $1/2$ at $v_{FT}=1.234$, and 
hence system enters two critical modes case. Further increase of sweep velocity raises also the transition probability at global minimum ($p_{k^{+}_m}$) 
and DQPTs are wiped out for $v> v_c=28.084$.

As clear, for a quench crossing three critical points where the end of quench does not limited between two critical points, the system reveals critical sweep velocity
above which DQPTs are disappeared. This behaviour is also observed for a quench that crossing two critical points for $J_3<1/2$, where the ramp field end does not confined 
between two critical point (\ref{QASTCP}-(i)).

\subsubsection{Dynamical phase diagram}
	 	
In Fig. \ref{Fig3}(f), the dynamical phase diagram of the model has been displayed in $v-J_{3}$ plane for $h_{f} = 1.5$.
The dynamical phase diagram represents dynamics of the system for three cases of quench: crosses two critical points where quench field end does not confined between two critical points ($J_3<1/2$), across two critical points where quench field end limited between two critical points ($1/2<J_3<5/2$) and passage through a single anisotropic transition ($J_3>5/2$). The phase diagram discloses five distinct regions, no-DQPTs, single critical modes (SCM), two critical modes (TwCMs) and three critical modes (ThCMs).
As seen, for $J_3<1/2$ where the quench crosses two Ising like transition points ($h^{(1)}_c, h^{(2)}_c$) and the end of quench is not restricted between two critical points ($h_f>h^{(2)}_c$), the system reveals critical sweep velocity (red solid line) above which DQPTs are disappeared. While for $v<v_{c}$ system experiences DQPTs with two critical modes. For $1/2<J_3<5/2$ although the quench passage through two critical points, the ramp filed end is confined between two critical points ($h^{(1)}_c<h_f<h^{(2)}_c$) and DQPTs always are present. In such a case, the system encloses three critical modes region which passes to single critical mode for $v>v_T$
(blue solid line). As $J_3$ exceeds $5/2$ the quench from $h_i=-10$ to $h_f=1.5$ crosses only a single anisotropic transition point ($h^{(a)}_c$) which displays two critical modes.

Fig. \ref{Fig4} represents the dynamical phase diagram of the model in the $v-J_{3}$ plane for the quench to $h_{f}=10$, which crosses two critical points for $J_3<1/2$ and passage through three critical points for $J_3>1/2$. As seen, the dynamical phase diagram contains three distinct regions: four critical modes (FCMs), two critical modes (TwCMs), and no-DQPTs regions. The red solid line in the dynamical phase diagram indicates the critical sweep velocity ($v_c$) which separates no-DQPTs region from two critical modes region. The blue dashed-dotted line represents $v_T$ under which the system enters four critical modes region which appears only for $J_3>1/2$ where quench pass over three critical points. 

From these observations, we come to the conclusion that, DQPTs always happen if the starting or ending point of the ramp field confined between two critical point, even for sudden quench case. While for a quench that starting or ending point of the ramp field does not restricted between two critical points, there exist critical sweep velocity 
above which DQPTs are wiped out.

%
\begin{figure*}[]
\begin{minipage}{\linewidth}
\centerline{\includegraphics[width=0.33\linewidth]{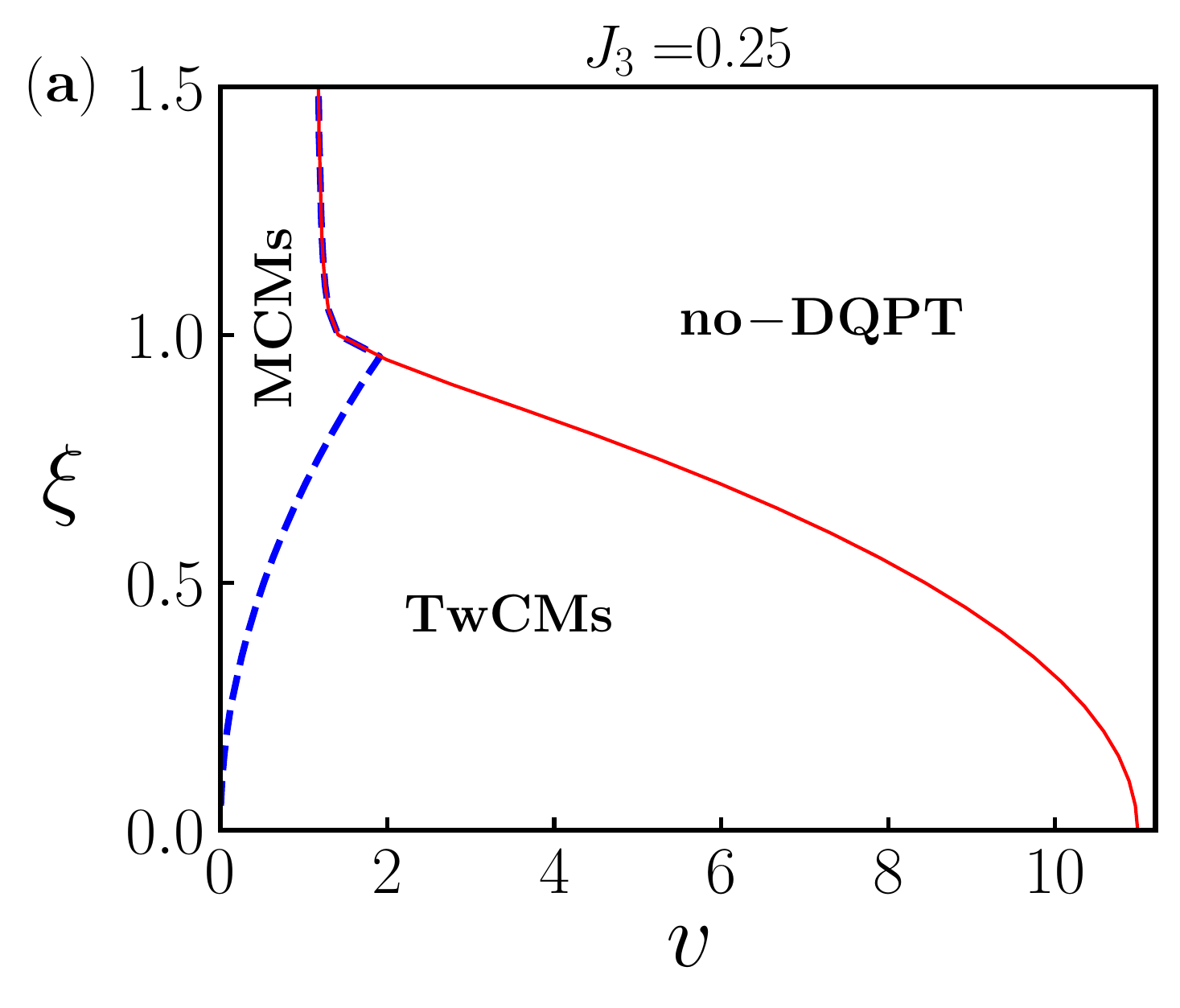}
\includegraphics[width=0.33\linewidth]{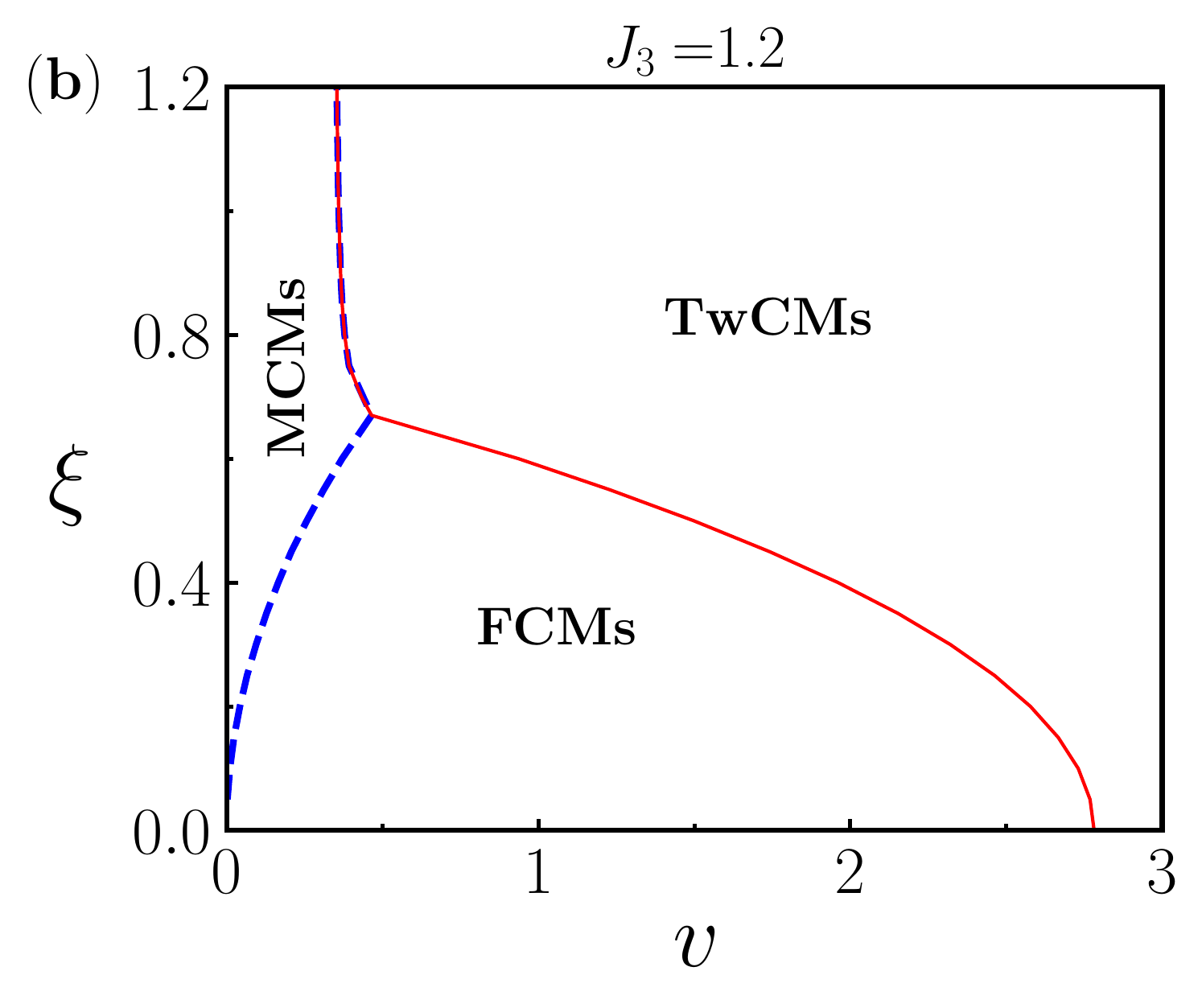}
\includegraphics[width=0.33\linewidth]{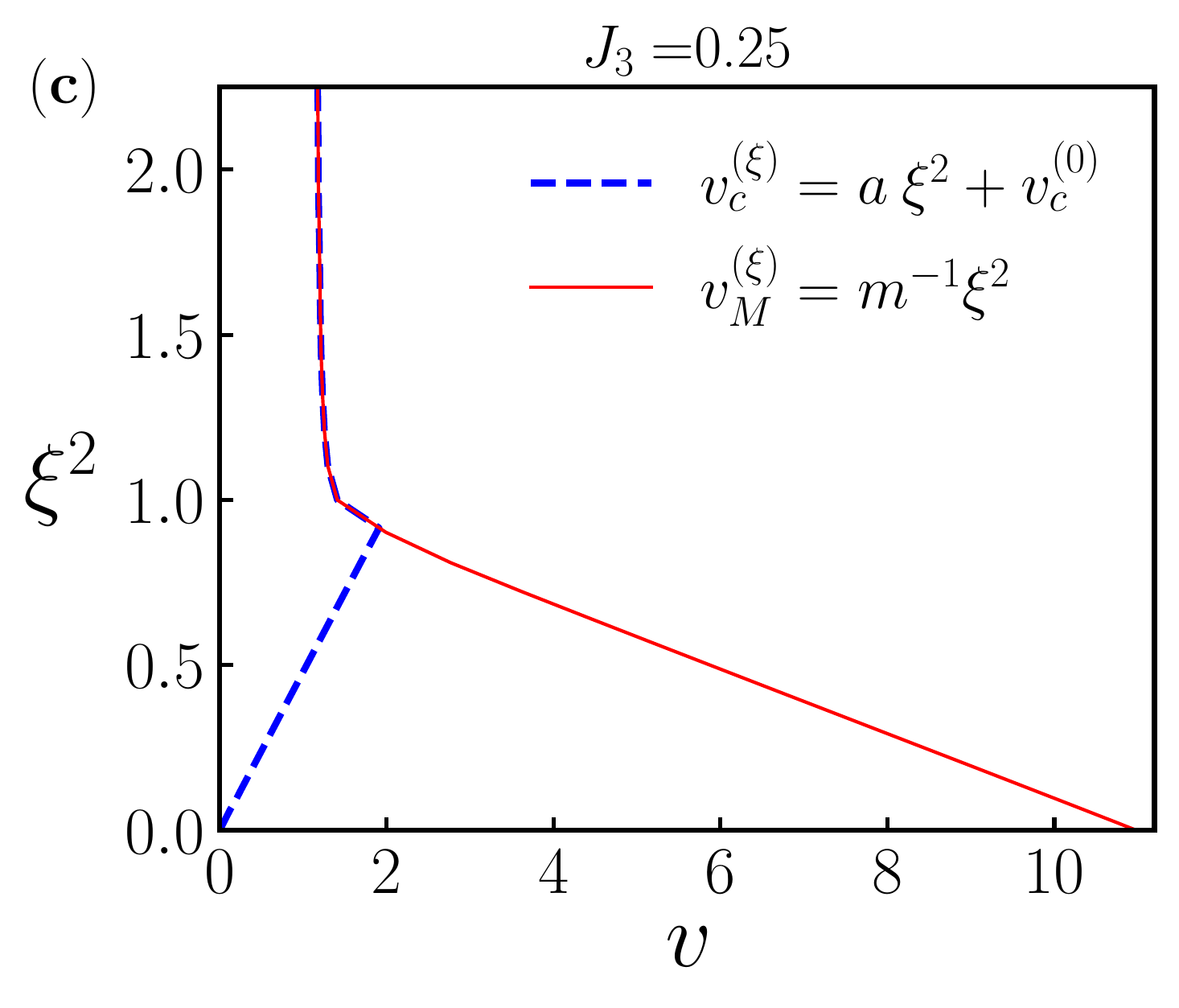}}
\centering
\end{minipage}
\begin{minipage}{\linewidth}
\centerline{\includegraphics[width=0.33\linewidth]{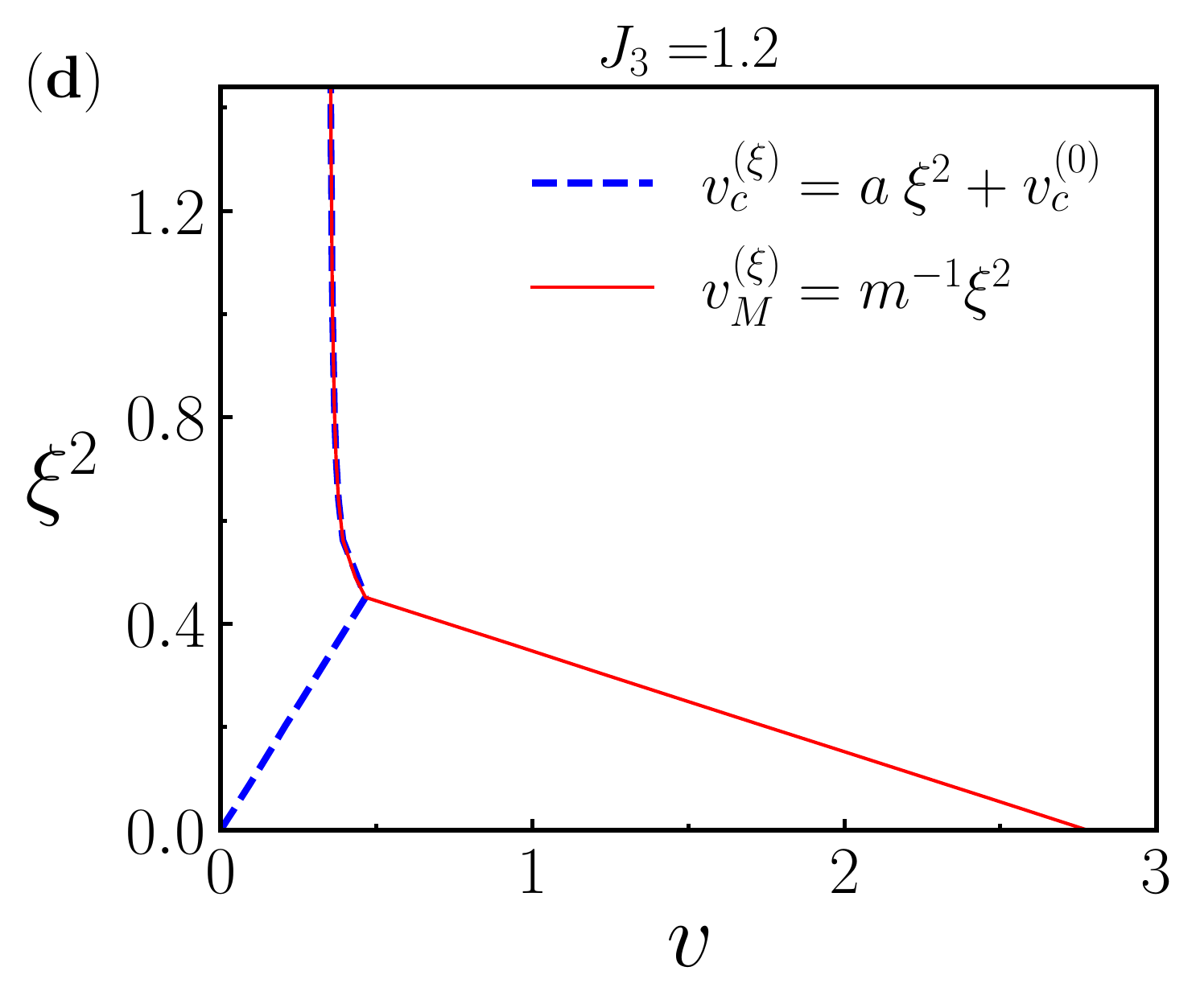}
\includegraphics[width=0.33\linewidth]{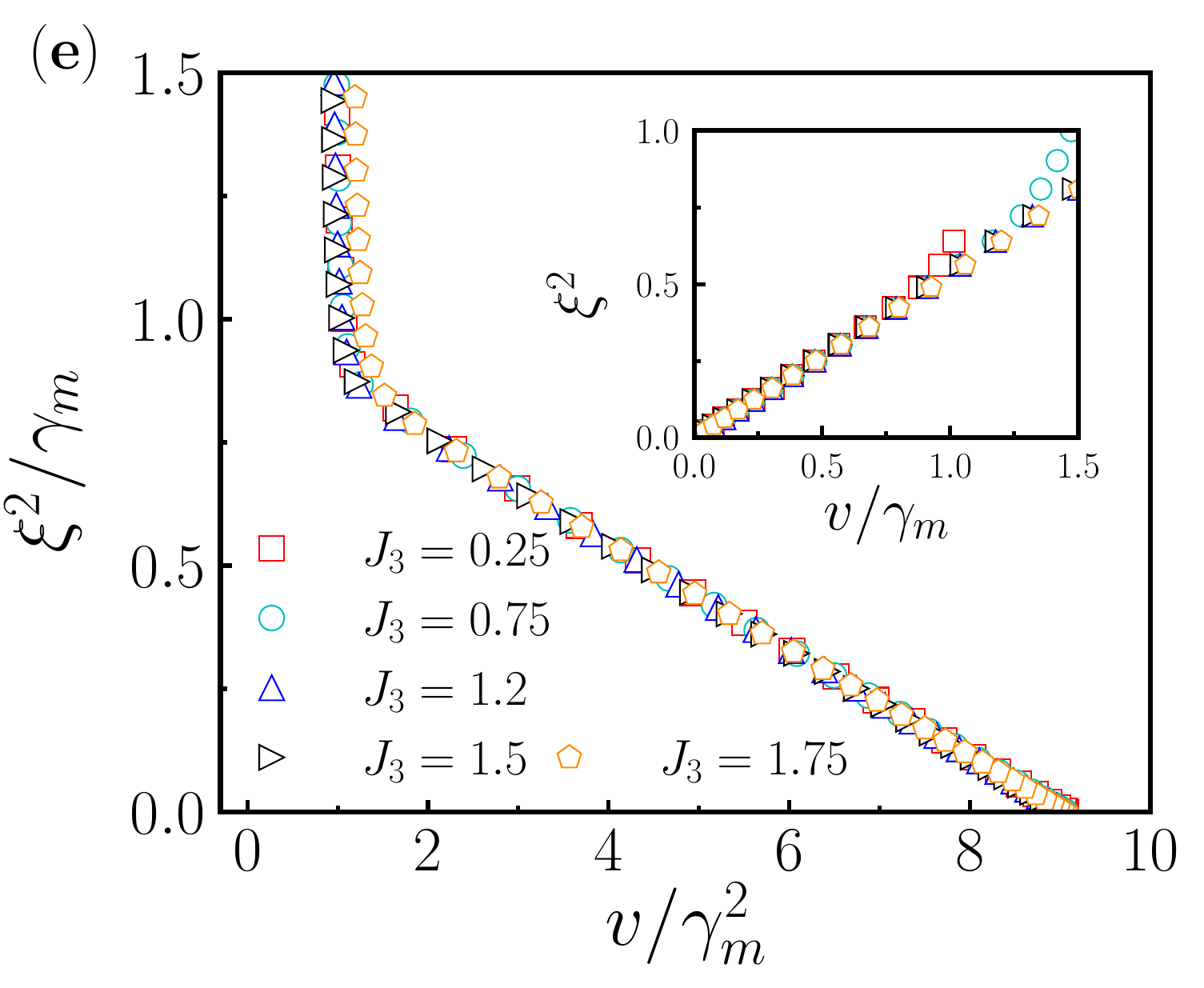}
\includegraphics[width=0.33\linewidth]{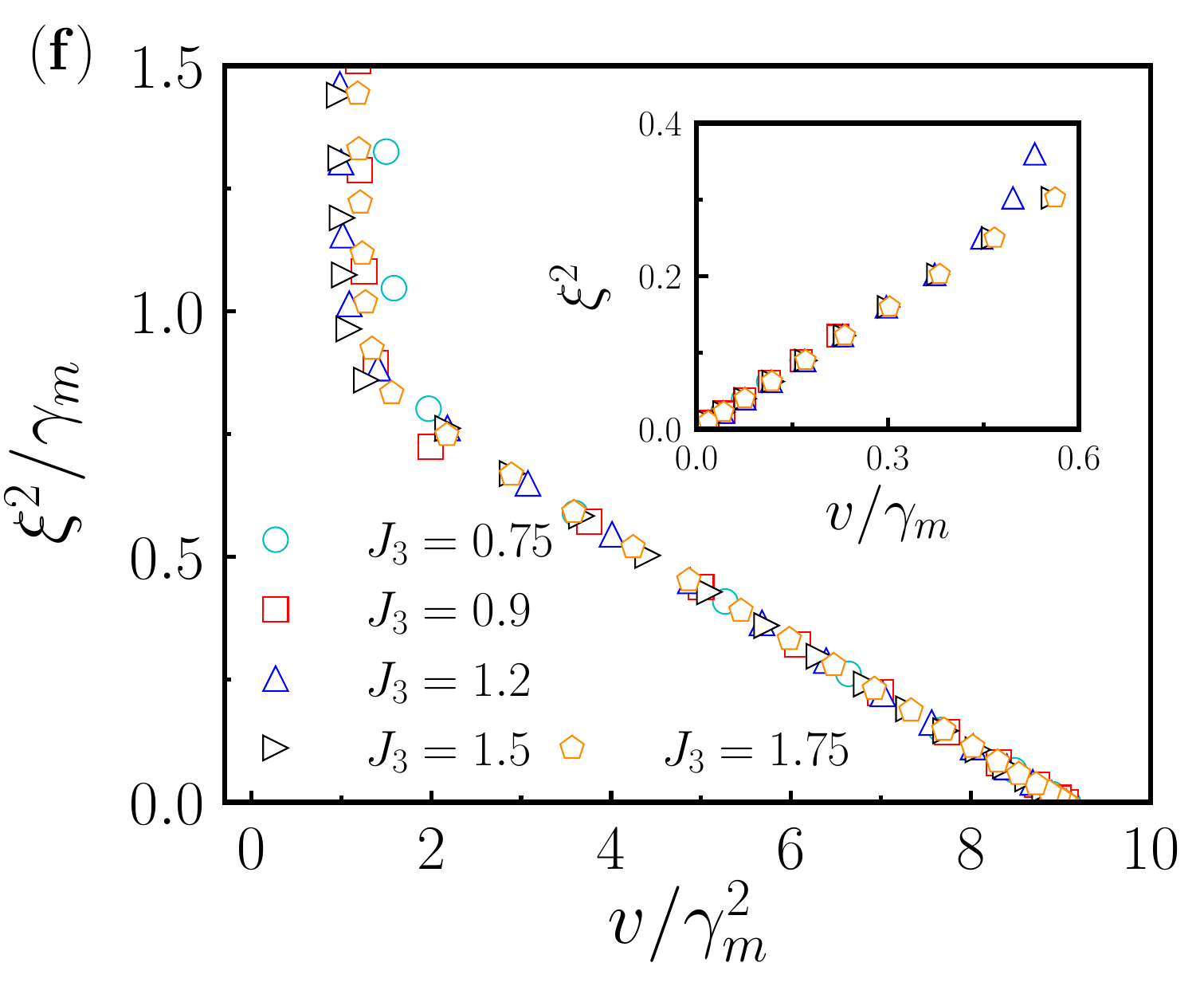}}
\centering
\end{minipage}
\caption{(Color plot) The phase diagram in the $v-\xi$ plane 
for (a) $ J_{3}=1/4$ and (b) $ J_{3}=1.2$. The phase diagram includes 
two regions: DQPTs, no-DQPT regions. DQPTs region is classified into two 
regions: multi-critical modes (MCMs) ($v<v_{M}$) region and two critical modes (TwCMs) 
region. The phase diagram in the $v-\xi^2$ plane for (c) $J_{3}=1/4$ and (d) $J_3=1.2$, represent linear scaling 
of $v_c$ and $v_M$ with $\xi^2$. (e) and (f) represent universal scaling function under which $v_c$ curves
corresponding to the different values of $J_3$ collapse on a single graph. 
Insets: the universal scaling function of $v_M$.}  
\label{Fig6}
\end{figure*}


\section{NOISY RAMP QUENCH}\label{section5}
	
As noted, noise is an unavoidable and inescapable in any physical system. Consequently, comprehending the exact effects of noise is crucial in both classical and quantum contexts \cite{Jafari2024,Baghran2024,Rahmani2016,Bando2020,Chenu2017,Gao2017,Ai2021,Singh2021}. Noise introduces randomness into the system, thereby disrupting coherent evolution. In this section, 
we investigate DQPTs in the one-dimensional quantum Ising model with cluster interaction, following a ramp protocol with fluctuations. To this end, we add uncorrelated Gaussian noise to the time dependent transverse field, expressed as $h(t) = h_{f} +vt+R(t)$, where  $R(t)$ represents a random fluctuation confined to the ramp interval $[t_{i}, t_{f }= 0[$, with a vanishing mean, $\langle R(t) \rangle = 0 $. We use white noise with Gaussian two-point correlations $\langle R(t)R(t') \rangle = \xi^{2} \delta (t-t')$ where $\xi$ characterizes the strength of the noise ( $\xi^2$ has units of time).
	
To determine the transition probability in the presence of noise, we solve numerically the exact noise master equation \cite{Jafari2024,Baghran2024,Luczka1991,Budini2000,Costa2017,Kiely2021} for the averaged density matrix $\rho_{k}(t)$ of the Hamiltonian which includes noise $H^{(\xi)}_{k} = H^{(0)}_{k} (t)+R(t)H_{1}$; where $H^{(0)}_{k} (t)$ denotes noiseless Hamiltonian and $R(t)H_{1}=2R(t)\sigma^{z}$ represents the added noise which appears during the ramp process where $t \in [t_{i}, t_{f }[$.  The exact noise master equation in the presence of the uncorrelated Gaussian noise is given as follows \cite{Jafari2024,Baghran2024,Rahmani2016,Bando2020,Gao2017}
%
{\small
\begin{equation}
\label{eq16}
\frac{d}{dt} \rho_{k}(t)=-i\Big[H^{(0)}_{k}(t),\rho_{k}(t)\Big]-\frac{\xi^{2}}{2}\Big[H_{1},\big[H_{1},\rho_{k}(t)\big]\Big]
\end{equation}
}
%
By numerically solving the master equation, the transition probability in the presence of uncorrelated Gaussian noise 
is given as \cite{Jafari2024,Baghran2024,Rahmani2016,Bando2020,Chenu2017}
%
\begin{equation}
\label{eq16}
p_{k}=\langle \beta_{k}(t_{f})| \rho_{k}(t) | \beta_{k}(t_{f}) \rangle .
\end{equation}
%
where, $| \beta_{k}(t_{f}) \rangle$ is the excited state of the system at the end of quench (Eq. \eqref{eq12}).
The dynamical phase diagram of the model is characterized by the interplay of the near-adiabatic dynamics of the system’s gapped fermionic modes and the accumulation of noise-induced excitations during the evolution.
Moreover, we expect that large values of the sweep velocity gives less time for the noise to be effective. 
In this section we have searched the effects of noise on transition probability and dynamical phase diagram of the model.


\subsection{Transition probability}
Without loss of generality and prevent complexity we suppose that in the presence of the noise $h(t)$ altering from $h_{i}=-10$ to $h_{f}=10$, which crosses all critical points. Therefore, the quench filed end is not blocked between two critical points and the system shows transition from DQPTs region to no-DQPTs region. 
From a physical point of view, the influence of noise on transition probability and critical sweep velocity is an interesting open question: 
Will noise reduce or enhance the critical sweep velocity or are DQPTs always possible?
particular, we wish to determine how the dynamical phase diagram of the model is modified in the presence of noise.

Fig. \ref{Fig5} illustrates the transition probability as a function of $k$ for different values of sweep velocities and noise intensity for $J_3=1/4$ and $J_3=1$. As observed, for $\xi/v \ll 1$ , the effect of noise is to displace the critical mode $k^*$, resulting in a sequence of DQPTs. The gradual increase in noise intensity, the effect of noise turns in an unanticipated direction for $ \xi/v \sim {\cal{O}}(1)$. In such a case, $p_k$ curve is locked to the value $1/2$ over a finite interval of momenta, leads to multi-critical modes (MCMs). This suggests that the noise behaves like a high-temperature source, leading to maximally mixed states unless the $k$-modes are too “light” (easily excited to the upper level by the Kibble-Zurek Mechanism \cite{Rahmani2016}).
	
Moreover, the main effect of noise during ramping time is that the inequality $p_{k_{m}}^{min} < 1/2$ fulfilled only for sufficiently low noise amplitudes. In other words, very strong noise ($\xi/v \gg 1$) leads to non-adiabatic transitions of such high probability that a maximally mixed state ($p_{k} = 1/2$) is not observed at the end of quench, consequently preventing the emergence of DQPTs even for $v < v_c$. Therefore, the boundary between ”DQPTs” and ”no-DQPTs” regions is changed in the presence of noise. In the following, we will explore the dynamical phase diagram of the model for different noise intensity to clarify the process of noise effects on DQPTs.


\subsection{Dynamical phase diagram and scaling of critical sweep velocity}
The phase diagram of the model in the presence of noise has been plotted in Fig. \ref{Fig6}(a)-(f) in $ v-\xi$ plane for $J_{3}=1/4$ and $J_{3}=1.2$. The numerical results indicate that, in the presence of noise the critical sweep velocity ($v^{(\xi)}_c$) above which the DQPTs disappear, decreases by enhancing the noise intensity $\xi$ (red solid line in Fig.\ref{Fig6}(a)). These findings are consistent with our expectation that the noise induces non-adiabatic transitions and a maximally mixed state ($p_{k}= 1/2$) does not occur at the end of quench, thus hindering the emergence of DQPTs.
	
As seen in the noiseless case for $J_{3}<1/2$, there are two critical modes for $v < v_{c}$ at which DQPTs occur. Since in the presence of noise, the $p_k$ curve locked to the value $1/2$ over a finite interval of momenta, the DQPTs region is divided into two regions: multi-critical modes and two critical modes regions. In Fig. \ref{Fig6}(a), the left corner of the phase diagram marked MCMs represents the multi-critical modes region, which is separated from the two critical modes (TwCMs) region by the dashed blue line.  

On the other hand, when $J_{3} > 1/2$, in the noiseless phase diagram (Fig. \ref{Fig4}) the DQPTs contains four critical modes region for $v < v_{FT}$ and two critical modes region for $v_{FT} < v < v_{c}$. In the presence of noise, the FCMs region splits into two regions: MCMs and FCMs regions. In Fig. \ref{Fig6}(b) the left corner of the phase diagram marked MCMs represents the multi-critical modes region which is separated from the four critical modes (FCMs) region by the blue dashed line.
	
As seen, the sweep velocity below which the system enters the MCMs region $v_M$ (blue dashed line) increases by enhancing the noise intensity while the critical sweep velocity $v_c$ (red solid line) reduces by increasing the noise. The numerical results indicate that the $v_M$ converges with the critical sweep velocity $v_c$ curve, and consequently, both curves merge into a single line for very strong noise ($ \xi > 1 $).

Having established that the dynamical phase diagram of the model is modified in the presence of the noise, the question then arises, does the system show any scaling and universality in the presence of noise? To this end, the critical sweep velocity $v_c$ and $v_M$ have been plotted in the $v-\xi^2$ plane, as shown in Fig. \ref{Fig6} (c)-(d).
	
The numerical analysis shows that, the critical sweep velocity scales linearly with the square of noise intensity for both weak ($\xi \sim {\cal O}(10^{-2})$) and strong ($\xi \sim {\cal O}(10^{-1})$) noise, i.e., $v_{c}^{(\xi)}=a \xi^{2}+v_{c}^{(0)}$ where $v_{c}^{(0)}$	represents the critical sweep velocity for the noiseless case.
Additionally, a similar linear scaling is observed for $v_{M}$ under both weak and strong noise conditions, given by $v_{M}^{(\xi)} = \xi^{2}/m$.
	
A more detailed analysis shows that the slope of lines depicted in Fig. \ref{Fig6}(c)-(d) also scales with $\gamma_{m}$ (see Appendix \ref{c}). The numerical results demonstrate that the slops $a$ and $m$ scale in a power law manner with $\gamma_{m}$ with exponent $\alpha$, i.e., $\{a, m\} \propto  \gamma_{m}^{|\alpha|}$ with $|\alpha| = 1 \pm 0.001 $ for $\{a, m\}$.
	
A more numerical examination for both weak and strong noise illustrate a potential extrapolation for a scaling behavior of $v_{c}^{(\xi)}$ i.e., $v_{c}^{(\xi)}$ is invariant under the scaling transformation $v_{c}^{(\xi)} \to v_{c}^{(\xi)}/\gamma_{m}^2$ and $\xi \to \xi/\sqrt{\gamma_{m}}$. Moreover, the scaling function for multi-critical sweep velocity is $v_{M} \to v_{M}/\gamma_{m}$.
	
The scaling of critical sweep velocity $v_{c}^{(\xi)}$ corresponding to different values of $J_{3}$, for both weak and strong noise, has been illustrated in Fig.  \ref{Fig6}(e)-(f) in which all curves collapse to a single graph under the scaling function. The scaling function for $v_M$ below which the system reveals multi-critical 
modes has been also displayed in the inset of Fig. \ref{Fig6}(e)-(f). These scaling functions are the promised universality of DQPTs in the presence of the noise.


\section{Conclusions}\label{section6}
	
We have studied the dynamical quantum phase transition (DQPTs) in the one dimensional Ising model with cluster interaction in the presence of the 
the linear time dependent transverse field. The usual symmetry in the equilibrium phase diagram of the transverse field Ising model is broken in the presence of the cluster 
interaction. In time-independent transverse field case, besides the two Ising like quantum phase transition points where gap closing occurs at high symmetry 
point in the Brillouin zone, there is a quantum phase transition between paramagnetic and cluster phases where the gap closing mode can be moved by tuning the cluster interaction strength. We have shown that, the modes which are confined between two gap closing modes (maximum transition probability ($p_k=1$)) can be easily excited to the upper level. While the transition probability of the farthest mode from the single gap closing mode remains zero even for a sudden quench case. 
Consequently, DQPTs always occur for a quench that starting or ending point of the quench is limited between two critical points. In other respects there is always a critical sweep velocity above which DQPts are wiped out. 
Moreover, our finding also confirmed in the presence of the noisy quench while the critical sweep velocity decrease in the presence of the noise.
In addition, a surprising result occurs when the noise intensity and sweep velocity are in the same order of magnitude where the transition probability
is locked to $1/2$ over a finite range of momentum. In such a case, the multi critical modes region and consequently multi-critical time scale is induced in 
dynamical phase diagram. The analysis shows that the critical sweep velocity above which DQPTs disappear scales linearly with the square of noise intensify.
Furthermore, the sweep velocity below which the system enters the multi-critical modes region, increase by enhancing the noise intensity and also scales linearly 
with the square of noise intensity.    
%
\begin{figure}[h!]
\begin{minipage}{\linewidth}
\centerline{\includegraphics[width=0.48\linewidth]{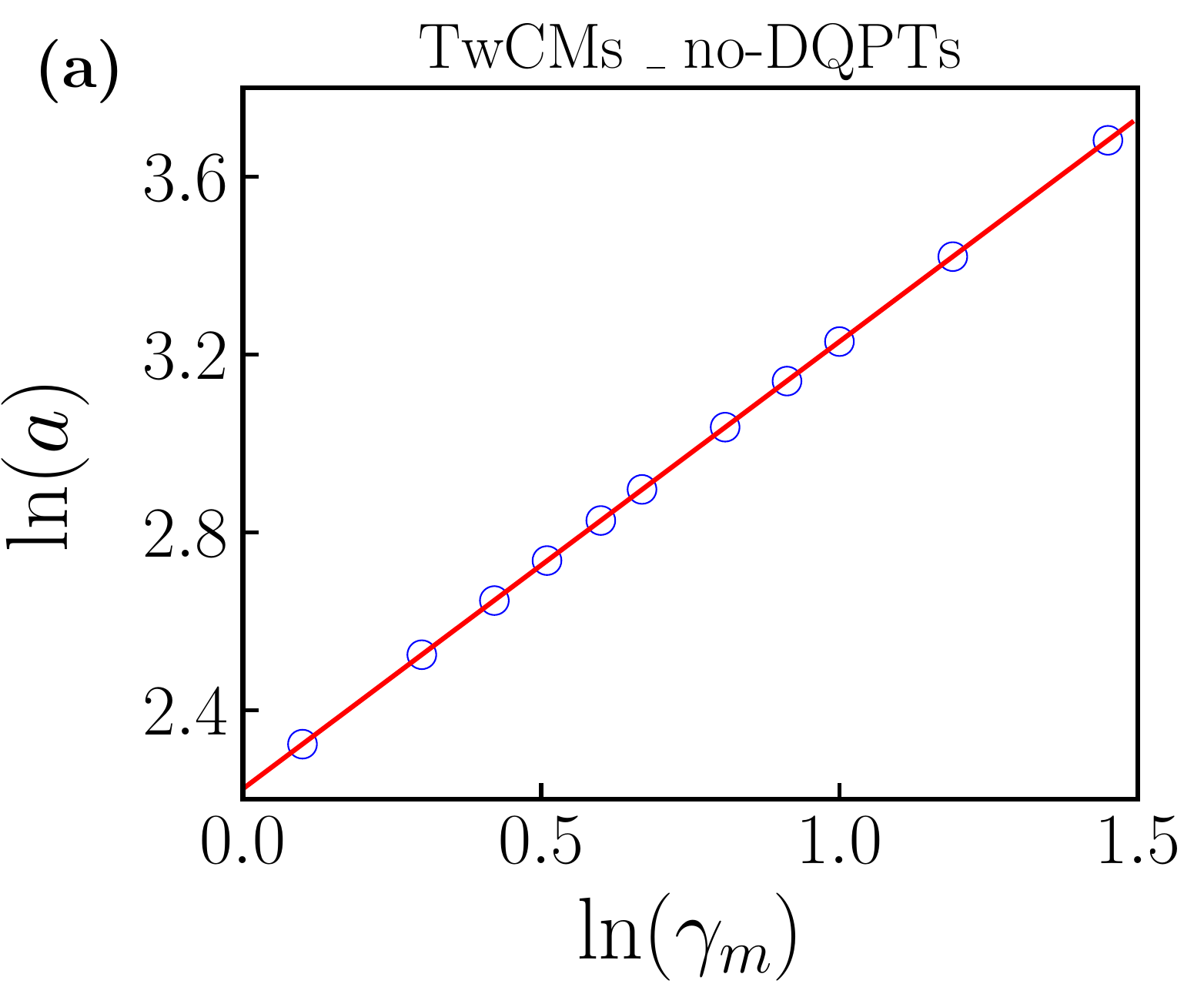}
\includegraphics[width=0.47\linewidth]{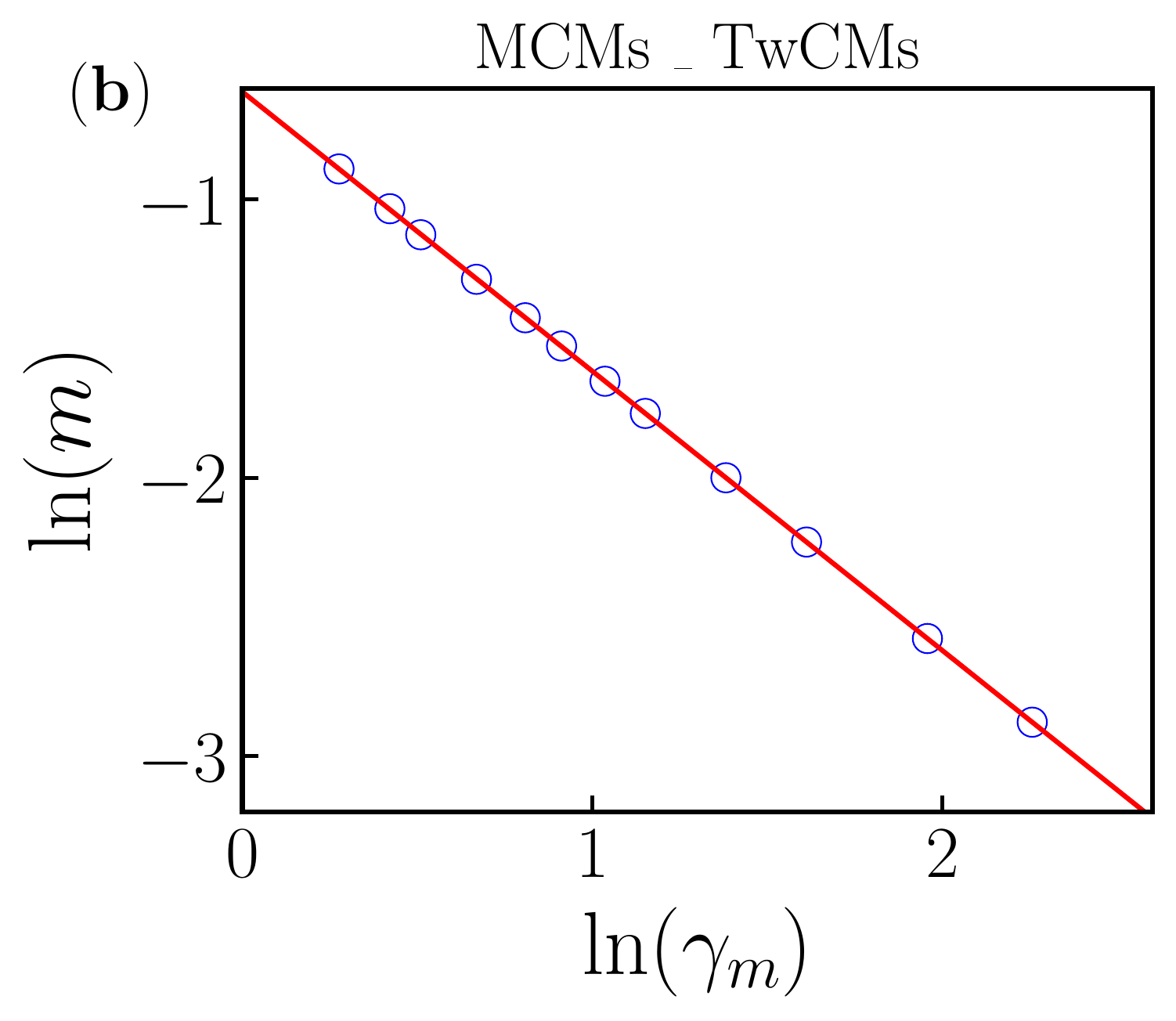}}
\centering
\end{minipage} 
\begin{minipage}{\linewidth}
\centerline{\includegraphics[width=0.48\linewidth]{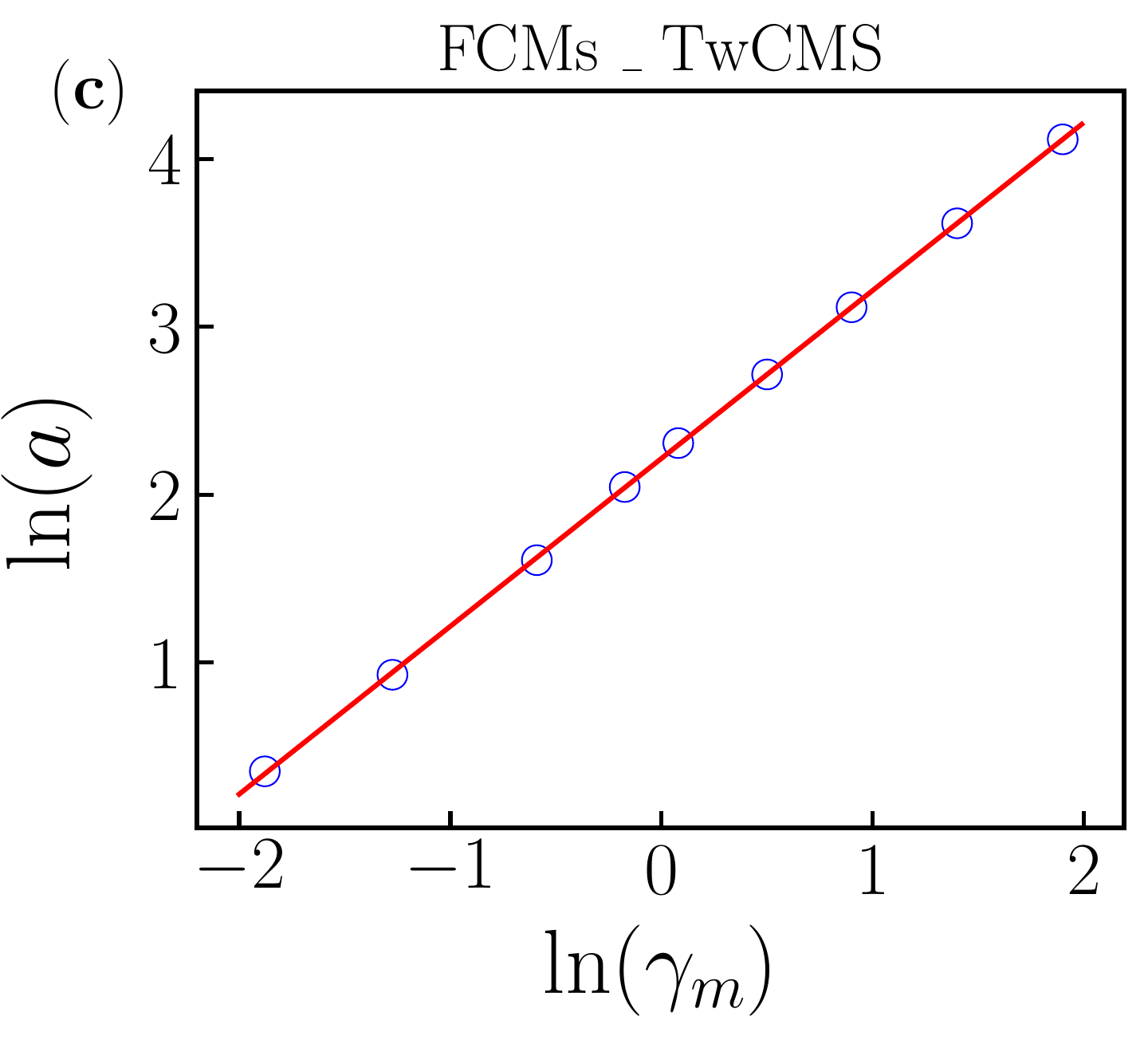}
\includegraphics[width=0.49\linewidth]{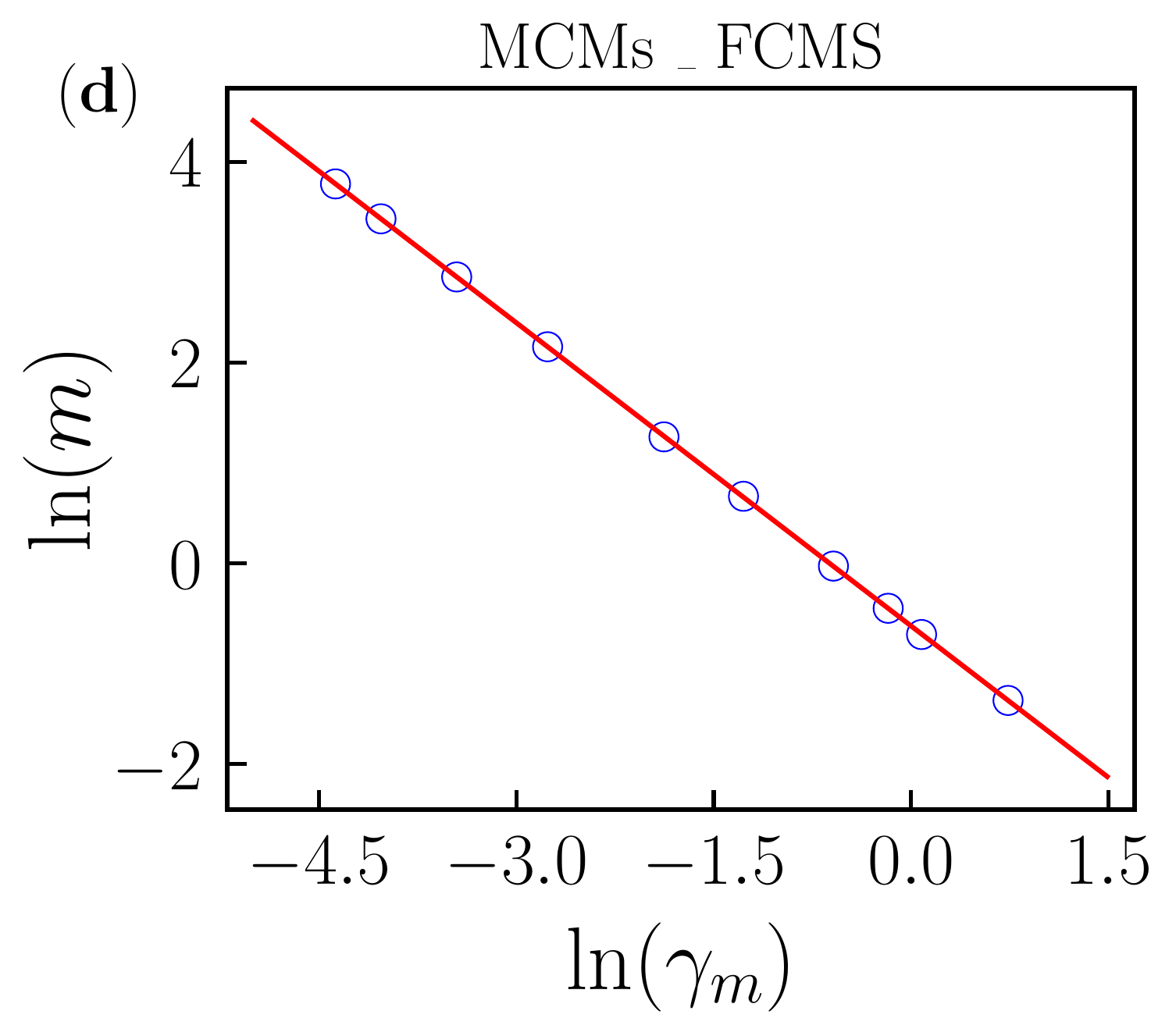}}
\centering
\end{minipage}
\caption{(Color plot) The scaling of the slope of linear lines corresponding to Fig. \ref{Fig6}(b,e) with $\gamma_{m}$}
\label{slope1}
\end{figure} 
%
	



\appendix
	
\section{Time dependent Sch\"odinger in the diabatic basis}\label{a}
	
The time-dependent Schr\"odinger equation associated with the Hamiltonian presented in Eq. \eqref{H2} can be expressed as follows
%
\begin{eqnarray}
i\frac{d}{dt}\begin{pmatrix}
a_{1}(t) \\
a_{2}(t) 
\end{pmatrix}=\begin{pmatrix}
A(k,t) & B(k) \\
B(k)&-A(k,t)
\end{pmatrix}\begin{pmatrix}
a_{1}(t) \\
a_{2}(t) 
\end{pmatrix}
\end{eqnarray}
%

where $A\left( {k,t} \right) = 2\left( {h\left( t \right) - {J}\cos k - {J_3}\cos 2k} \right)$ and $B\left( k \right) = 2\left( {{J}\sin k + {J_3}\sin 2k} \right)$
and $a_{1}(t)$, $a_{2}(t)$ are the coefficients that characterize the quantum state in the diabatic basis.
	 
By introducing new variables as follows  
%
\begin{eqnarray}
\tau &=&2(vt-J\cos k- J_{3}\cos 2k)/v \\[5pt]
\gamma &=&  {{J}\sin k + {J_3}\sin 2k} \nonumber
\end{eqnarray}	
%
the Hamiltonian
%
\begin{eqnarray}
H_{k}(t) = \begin{pmatrix}
A(k,t) & B(k) \\
B(k)&-A(k,t)
\end{pmatrix}
\end{eqnarray}
%
can be mapped to the Landau-Zener counterpart $H_{LZ}(\tau)$	
%
\begin{eqnarray}
H_{LZ}(\tau)= \begin{pmatrix}
v\tau & \gamma \\
\gamma&-v\tau 
\end{pmatrix}.
\end{eqnarray} 
%
This can be resolved analytically\cite{Vitanov1996,Vitanov1999}.	 

\section{Scaling of the slope}\label{c}

Fig. \ref{slope1} illustrates the relationship between the slope of the linear lines presented in Fig. \ref{Fig6}(c,d) and the variable $\gamma_{m}$. It is evident that, in all instances, the slope of the lines exhibits a linear dependence on $\gamma_{m}$.

\bibliography{Ref-TFIM-Cluster}
	
\end{document}